\documentclass[paper=a4,abstract=true,numbers=noenddot]{scrartcl}

\usepackage[english]{babel}  
\usepackage{hyperref}
\usepackage[intlimits]{amsmath,nccmath}
\usepackage{amsthm}
\usepackage{amssymb} 
\usepackage{units} 
\usepackage{amsfonts,mathrsfs,bbm}
\usepackage{caption} 
\usepackage{booktabs}
\allowdisplaybreaks[1]
\usepackage[numbers]{natbib}
\usepackage{graphicx,subcaption}
\usepackage[space]{grffile} 
\usepackage{tikz}
\usetikzlibrary{calc}
\usepackage{pgfplots}
\pgfplotsset{compat=1.18} 
\pgfplotsset{every axis/.append style={thick}}
\pgfplotsset{every axis legend/.append
  style={cells={anchor=west},anchor=west}}
\pgfplotscreateplotcyclelist{mylines}{%
  {red,mark=*},
  {blue,mark=square},
  {green,mark=+},
  {black,mark=o},
  {violet,mark=otimes},
  {brown,mark=triangle},
  {cyan,mark=diamond},
  {orange,mark=10-pointed star},
  {magenta,mark=|}  }
\usepackage[T1]{fontenc}
\usepackage{mathptmx}
\usepackage[scaled=.92]{helvet}
\usepackage{courier}

\usepackage[a4paper,left=2.5cm,right=2.5cm,top=3cm,bottom=2.5cm]{geometry}
\usepackage{authblk}
\usepackage{amsmath,bm}
\usepackage{upgreek}
\usepackage{mathtools}
\usepackage{xcolor}
\usepackage{algorithm}
\usepackage{algpseudocode}
\usepackage{algorithmicx}
\theoremstyle{remark}

\usepackage{stmaryrd}
\usepackage{cleveref}
\usepackage{mymacros}

\title{A Higher-Order Time Domain Boundary Element Formulation based
  on Isogeometric Analysis and the Convolution Quadrature Method} 

\author[1]{Thomas Kramer}
\author[1]{Benjamin Marussig}
\author[1]{Martin Schanz}
\affil[1]{Institute of Applied Mechanics, Graz University of
  Technology, Technikerstraße 4/II, 8010 Graz, Austria,
  kramer@tugraz.at, marussig@tugraz.at, m.schanz@tugraz.at}
\date{}                     
\graphicspath{{pics/}{pics_tmp}}
\begin{document}
	
\maketitle
\section*{Abstract}
An isogeometric boundary element method (BEM) is presented to solve scattering problems in an isotropic homogeneous medium. We consider wave problems governed by the scalar wave equation as in acoustics and the Lamé-Navier equations for elastodynamics considering the theory of linear elasticity. The underlying boundary integral equations imply time-dependent convolution integrals and allow us to determine the sought quantities in the bounded interior or the unbounded exterior after solving for the unknown Cauchy data. 
In the present work, the time-dependent convolution integrals are approximated by multi-stage Runge-Kutta (RK) based convolution quadratures that involve steady-state solutions in the Laplace domain. 
The proposed method discretizes the spatial variables in the framework of isogeometric analysis (IGA), entailing a patchwise smooth spline basis. Overall, it enables high convergence rates in space and time. 
The implementation scheme follows an element structure defined by the non-empty knot spans in the knot vectors and local, uniform Bernstein polynomials as basis functions. 
The algorithms to localize the basis functions on the elements are outlined and explained. The solutions of the mixed problems are approximated by the BEM based on a symmetric Galerkin variational formulation and a collocation method. We investigate convergence rates of the approximative solutions in a mixed space and time error norm.

\textbf{Keywords}: wave equation; boundary element method; convolution quadrature; isogeometric analysis

\section{Introduction}
Wave propagation problems appear often in engineering, e.g.,~the sound radiation of machines, non-destructive testing or exploring the underground. Most of these problems are formulated in terms of hyperbolic partial differential equations, such as those governing acoustics or elastodynamics. The handling of space and time requires expensive discretization methods, where for scattering problems even an unbounded domain has to be considered. The latter are elegantly solved with the boundary element method (BEM). Its basis is
boundary integral equations with retarded potentials as kernels, which build the counterpart to the governing hyperbolic partial differential equation. The mathematical theory goes back to the beginning of the last century
by Fredholm for scalar problems like acoustics and later by Kupradze~\cite{kupradze79} for vectorial problems in elasticity. The mathematical background of time-dependent boundary integral equations is summarized by Costabel~\cite{costabel04} and extensively discussed in the textbook by Sayas~\cite{sayas2016}.

Time-dependent problems can be treated with the BEM by direct computations in the time domain or by computations in the Laplace domain combined with an inverse transformation. 
Regarding the first, \textit{classical} time-stepping methods capture the time evolution based on tensor product basis functions in space and time and an analytical time integration~\cite{mansur1983time}. However, a stable time-stepping method relies on an adequate choice of the time step size \cite{wang2007concerning,peirce1997stability}. More stable time domain computations, so called, space-time methods, rely on an energetic boundary integral formulation and generalize Mansur's approach  considering time as a component in a $\mydim+1$ dimensional domain where $\mydim\in\Nat$ denotes the number of spatial dimensions. 
The space-time BEM has been studied recently \cite{polz2021space} and extended to IGA \cite{aimi2022iga}. This method allows adaptive meshes in space and time, however, computations in this high dimensional domain imply fast growing system matrices, i.e.,~for larger problems solving the equation system becomes tedious.

On the other hand, computations in the Laplace domain facilitate the treatment of time-dependent problems with the BEM as the hyperbolic equations of motion become elliptic partial differential equations \cite{cruse1968direct}. 
The time discretization method used in this work, the convolution quadrature method (CQM), proposed by Lubich \cite{lubich1988convolution}, can be seen as somewhat in between the direct time domain and Laplace domain methods. It is a quadrature rule that approximates the time domain convolution integrals with quadrature weights based on elliptic problems in the Laplace domain. The CQM does not only enhance a more stable time stepping procedure but also facilitates applications to visco- or poroelasticity since only the Laplace domain fundamental solutions are required \cite{schanz1997new, schanz2001application}. A multi-stage RK based CQM makes high convergence rates in time possible \cite{banjai2012runge}. Combined with IGA for discretizing the spatial variable, we obtain a higher-order discretization scheme in space and time. The IGA-BEM is well established \cite{beer2020isogeometric, falini2022collocation}. It is, not least, a good fit for simulations on geometries where solely boundary representations are available, as commonly found in CAD. In this work, IGA enables a discretization of the Laplace domain operators by continuous and discontinuous B-spline basis functions and evaluations of the integrants on an exact parametrization of the geometry. Frequency domain IGA-BEM has been studied recently \cite{wolf2020analysis, dolz2018fast, degli2023iga}. However, to the best of the authors' knowledge, this work presents a novel higher-order method in space and time through the combination of higher-order time domain approximation (RK-CQM) with higher-order Laplace-domain approximation (IGA-BEM). Regarding the implementation, the Bézier extraction \cite{borden2011isogeometric} allows us to assemble the matrices element-wise by representing the B-spline basis in terms of Bernstein polynomials. This basis transformation, together with a degree elevation algorithm \cite{piegl2012nurbs}, significantly simplifies the incorporation of smooth spline parameterizations into routines for Lagrange elements.

In \Cref{sec.problem}, we introduce the considered problems. \Cref{sec.discretization} outlines the discretization scheme where the time discretization via the CQM is explained in \Cref{subsec.time_discretization} and the isogeometric discretization in the spatial variable is presented in \Cref{subsec.space_discretization}. The latter includes notes on the needed algorithms to implement IGA into existing BEM codes. We discuss the expected convergence rates in space and time in \Cref{sec.optimal_convergence}, and validate them through numerical studies in \Cref{sec.examples}.

\section{Problem statement} \label{sec.problem}
Let the considered spatial domain be a bounded, simply connected Lipschitz domain $\domain\subset\Rn{\mydimension}$ with a boundary $\boundary=\partial\domain$. The outward unit normal is denoted by $\nvec{\xb}$ and the time variable is $\timevar\in[0,\infty)$. The two considered problem classes are acoustics and elastodynamics.

The scalar wave equation for a non-viscous fluid is given by
\begin{equation}\label{eq.acoustics1}
  \wavespeed^2\nabla^2\pressure(\xd,\timevar)=\frac{\partial^2\pressure}{\partial\timevar^2}(\xd,\timevar) \qquad (\xd,\timevar)\in\domain\times[0,\infty)\; ,
\end{equation} 
where the wave speed $\wavespeed=\sqrt{\kompression/\matdensity}$ is defined by the compression module $\kompression$ and the material density $\matdensity$. As usual $\nabla$ is the Nabla operator and $\nabla^2$ the Laplace operator. In the case of mixed problems, we decompose the boundary $\boundary:=\diriboundary\cup\neumboundary$ into non-overlapping Dirichlet and Neumann parts and apply the boundary conditions 
\begin{subequations}
\begin{align}
  \diritrace{\xd} \pressure(\xd,\timevar) & :=\lim_{\domain\ni\xd\to \xb\in\boundary}\pressure(\xd,\timevar)  = {\timefunction}_{\dirisub}(\xb,\timevar) &&  (\xb,\timevar)\in\diriboundary\times[0,\infty)\; , \\
  \neumtrace{\xd} \pressure(\xd,\timevar) & :=\lim_{\domain\ni\xd\to \xb\in\boundary}[\nvec{\xb}\cdot\nabla \pressure(\xd,\timevar)]   = {\timefunction}_{\neumsub}(\xb,\timevar) && (\xb,\timevar)\in\neumboundary\times[0,\infty)\; ,
\end{align}
\end{subequations}
where $\diritrace{\xd}$ and $\neumtrace{\xd}$ denote the respective Dirichlet and Neumann traces. The initial boundary value problem is complete by the homogeneous initial conditions $\pressure(\xd,0) = \frac{\partial\pressure}{\partial\timevar}(\xd,0) = 0$.  Hence, we solve for the unknown scalar pressure field $\pressure(\xd,\timevar)$ and the unknown flux $\flux(\xd,\timevar)=\nvec{\xb}\cdot\nabla\pressure(\xd,\timevar)$ with $\xb\in\boundary$.

Considering Hooke's law and a linear stress strain relation, the Cauchy stress tensor is given by
\begin{equation}
  \stresstensor(\xd,\timevar)=(\shearmodul(\nabla\displacement+\transpose{(\nabla\displacement)}) + (\kompression-\frac{2}{3}\shearmodul)\nabla\cdot\displacement\eye)(\xd,\timevar) \; .
\end{equation}
Based on this constitutive relation for elastodynamics, we obtain the vector-valued wave equation 
\begin{equation}\label{eq.elastodynamics1}
\kwavespeed^2\nabla\nabla\cdot\displacement(\xd,\timevar) - \swavespeed^2\nabla\times\nabla\times\displacement(\xd,\timevar) = \frac{\partial^2\displacement}{\partial\timevar^2}(\xd,\timevar)  \qquad (\xd,\timevar)\in\domain\times[0,\infty)\; ,
\end{equation}
where $\kwavespeed = \sqrt{(\kompression+\frac{4}{3}\shearmodul)/\matdensity}$ and $\swavespeed=\sqrt{\shearmodul/\matdensity}$ are the respective compression and shear wave speeds with  the shear modulus $\shearmodul$. The boundary conditions are 
\begin{subequations}
\begin{align}
  \diritrace{\xd} \displacement(\xd,\timevar) & :=\lim_{\domain\ni\xd\to \xb\in\boundary}\displacement(\xd,\timevar)  =  \diridata(\xb,\timevar) &&  (\xb,\timevar)\in\diriboundary\times[0,\infty)\; , \\
  \neumtrace{\xd} \displacement(\xd,\timevar) & :=\lim_{\domain\ni\xd\to \xb\in\boundary}[\nvec{\xb}\cdot\stresstensor(\xd,\timevar)]  = \neumdata(\xb,\timevar) &&  (\xb,\timevar)\in\neumboundary\times[0,\infty)\; ,
\end{align}
\end{subequations}
based on the Dirichlet and Neumann traces, respectively. Note that the Dirichlet trace is the same limiting process as in the acoustic case but in the Neumann trace the gradient of the Dirichlet data has to be replaced by the Cauchy stress tensor. Again, vanishing initial conditions $\displacement(\xd,0) = \frac{\partial \displacement}{\partial\timevar}(\xd,0) = \myvec{0}$ are assumed. Hence in the elastodynamic case, we solve for the unknown displacement field $\displacement(\xd,\timevar)$ and the unknown  traction vector field $\stress(\xd,\timevar)=\nvec{\xb}\cdot\stresstensor(\xd,\timevar)$ with $\xb\in\boundary$.

\paragraph{Boundary integral equations}
The solution to the problems stated above can be expressed via boundary integral equations based on the time domain fundamental solution $\fundsol(\xb-\yb,\timevar-\tau)$. The time-dependent fundamental solution does not depend on the absolute time $\timevar$ but on the time difference $\timevar-\tau$ with the variable $\tau\in(0,\timevar)$. Similarly, the fundamental solution depends on the distance $\dist=\|\xb-\yb\|$ rather than the spatial coordinates themselves. Without losing generality, the following derivations focus on the scalar wave equation, but also hold for the vector-valued case with the correlated fundamental solution and the respective vector-valued solutions. The representation formula in time domain is
\begin{align} \label{eq.timesol1}
  \pressure(\xb,\timevar)=  \int\limits_{0}^{\timevar}\int\limits_{\boundary}^{}\left[\fundsol(\xb-\yb,\timevar-\tau)(\neumtrace{\yd}\pressure)(\yb,\tau) - (\neumtrace{\yd}\fundsol)(\xb-\yb,\timevar-\tau)\pressure(\yb,\tau)\right]d\boundary_{\yb}d\tau\; ,
\end{align}
 for all $(\xb,\timevar)\in\domain \times[0,\infty)$. We apply the Dirichlet trace to \eqref{eq.timesol1} and exclude a ball $B_{\epsilon}(\xb):=\{\yd\in\domain:\abs{\yd - \xb}<\epsilon\}$ located at the singularity $\xb$ from the domain of integration. The limiting process $\epsilon\to 0$ breaks the integal term over the region enclosed by $B_{\epsilon}(\xb)$ into the difference of an countour integral over the boundary $\partial B_{\epsilon}(\xb)$, which holds the identity operator $\idty$, and an integral over $\partial B_\epsilon (\xb)\cap\domain$. We denote this difference as the time independent integral free term $\jump$. Defining the convolution of two functions by $(f*g)(\timevar)=\int_{0}^{\timevar}f(\timevar-\tau)g(\tau)d\tau$ the boundary integral equation for the collocation scheme reads
\begin{equation}\label{eq.collo_bie}
    \slp*\flux -  \dlp*\pressure = \jump{\timefunction}_{\dirisub}+\dlp*{\timefunction}_{\dirisub} - \slp*{\timefunction}_{\neumsub}\qquad (\xb,\timevar)\in\boundary\times[0,\infty)\; ,
  \end{equation}
with the respective time domain single and double layer integral operators
\begin{subequations} \label{eq.operators1}
  \begin{align}
    (\slp*\flux)(\xb,\timevar) & := \int\limits_{0}^{\timevar}\int\limits_{\boundary}^{}\fundsol(\xb-\yb,\timevar-\tau)\flux(\yb,\tau)d\boundary_\yb d\tau \; , \label{eq.slp} \\
    (\dlp*\pressure)(\xb,\timevar) & := \lim\limits_{\epsilon\to 0}^{}\int\limits_{0}^{\timevar}\int\limits_{\boundary\backslash B_\epsilon (\xb)}^{}\transpose{(\neumtrace{\yb}\fundsol)}(\xb-\yb,\timevar-\tau)\pressure(\yb,\tau)d\boundary_\yb d\tau\; .\label{eq.dlp}
  \end{align}
\end{subequations}
  To derive a Galerkin type boundary element formulation, we apply the traction or flux operator on \eqref{eq.timesol1} which results in a representation formula for the stress/flux. The set of equations can be established based on both representation formula. Applying the Dirichlet trace on \eqref{eq.timesol1} at the Dirichlet boundary and the Neumann trace at the Neumann boundary results in the symmetric set of equations
\begin{subequations} \label{eq:bie}
  \begin{align}
    \slp*\flux -  \dlp*\pressure &= \frac{1}{2}{\timefunction}_{\dirisub}+\dlp*{\timefunction}_{\dirisub} - \slp*{\timefunction}_{\neumsub} &(\xb,\timevar)\in\diriboundary\times[0,\infty)\; ,\label{eq.bie0}\\
    \adlp*\flux +  \hsop*\pressure &= \frac{1}{2}{\timefunction}_{\neumsub}-\adlp*{\timefunction}_{\neumsub} - \hsop*{\timefunction}_{\dirisub} &(\xb,\timevar)\in\neumboundary\times[0,\infty)\; ,\label{eq.bie1}
  \end{align}
\end{subequations}
with the additional time domain integral operators 
\begin{subequations} \label{eq.operators2}
  \begin{align}
    (\adlp*\flux)(\xb,\timevar) &:= \lim\limits_{\epsilon\to 0}^{}\int\limits_{0}^{\timevar}\int\limits_{\boundary\backslash B_\epsilon (\xb)}^{}(\neumtrace{\xb}\fundsol)(\xb-\yb,\timevar-\tau)\flux(\yb,\tau)d\boundary_\yb d\tau\; , \label{eq.adlp}\\
    (\hsop*\pressure)(\xb,\timevar) & := -\lim\limits_{\epsilon\to 0}^{}\int\limits_{0}^{\timevar}\neumtrace{\xb}\int\limits_{\boundary\backslash B_\epsilon (\xb)}^{}\transpose{(\neumtrace{\yb}\fundsol)}(\xb-\yb,\timevar-\tau)\pressure(\yb,\tau)d\boundary_\yb d\tau\; ,  \label{eq.hso}
  \end{align}
\end{subequations}
known as the adjoint double layer, and hyper singular operator, respectively. In case of the Galerkin formulation the singular points $\xb$ are located at quadrature points distributed over the smooth parts of the geometry. Hence, the integral free term in \eqref{eq:bie} is $\frac{1}{2}$.

  The single layer operator \eqref{eq.slp} is weakly singular. The remaining integrals involve differentiations with respect to their kernel functions, which increase the order of singularity. For acoustics, the double layer operator \eqref{eq.dlp} and its adjoint \eqref{eq.adlp} can be defined as an improper integral as the singularity can be lifted due to the orthogonality of the distance vector $\yb-\xb$ and the outward normal vector $\nvec{\yb}$. In elastodynamics,  those integrals have to be defined as Cauchy principal value integrals. The hyper singular operator \eqref{eq.hso} has to be understood as a finite part in the sense of Hadamard. To regularize the hyper singular and Cauchy singular integrals integration by parts is applied \cite{kielhorn2008convolution}. After this step, the resulting integral equations are weakly singular, which can be handled well in the discrete setting with the Duffy transformation~\cite{duffy2004transform}.

Alternatively to the formulations \eqref{eq.collo_bie} and \eqref{eq:bie}, which is often denoted as the direct formulation, the so-called indirect formulation based on the layer potentials can be used. Such kind of integral formulations are a good choice to test implementations since all the four operators in \eqref{eq.operators1} and \eqref{eq.operators2} can be tested separately. Based on the respective single and double layer potential
\begin{subequations}
\begin{align} \label{eq:single_layer}
  \pressure(\xb,\timevar) &=  \int\limits_{0}^{\timevar}\int\limits_{\boundary}^{} \fundsol(\xb-\yb,\timevar-\tau)\densityfirst(\yb,\tau) d\boundary_{\yb}d\tau
  && \forall (\xb,\timevar)\in\domain \times[0,\infty)\; ,   \\ \label{eq:double_layer} 
  \pressure(\xb,\timevar) &=   \int\limits_{0}^{\timevar}\int\limits_{\boundary}^{} (\neumtrace{\yd}\fundsol)(\xb-\yb,\timevar-\tau) \densitysecond(\yb,\tau) d\boundary_{\yb}d\tau
        &&\forall (\xb,\timevar)\in\domain \times[0,\infty)  \; ,
\end{align}
\end{subequations}
which are solutions of the underlaying partial differential equation, the indirect method can be established. The unknown density functions $\densityfirst(\xb,t)$ and $\densitysecond(\xb,t)$ have to be determined from the given boundary data. The respective integral formulations are for pure Dirichlet problems, i.e., $\neumboundary = \emptyset$
\begin{subequations} \label{eq:diri_indirect}
\begin{align}  \label{eq:slp_indirect}
  \slp*\densityfirst  &= \diribc \quad  (\xb,\timevar) \in\diriboundary\times[0,\infty)\; ,\\  \label{eq:dlp_indirect}
  \alpha \frac{1}{2} \densitysecond + \dlp *\densitysecond & = \diribc \quad
                                   (\xb,\timevar)\in\diriboundary\times[0,\infty)\; ,
\end{align}
\end{subequations}
and for pure Neumann problem, i.e., $\diriboundary = \emptyset$
\begin{subequations} \label{eq:neum_indirect}
\begin{align}  \label{eq:adlp_indirect}
  - \left(\alpha \frac{1}{2} \densityfirst - \adlp *\densityfirst \right)& = \neumbc \quad  (\xb,\timevar) \in\neumboundary\times[0,\infty)\; ,\\  \label{eq:hyper_indirect}
  - \hsop*\densitysecond  &= \neumbc \quad
                  (\xb,\timevar)\in\neumboundary\times[0,\infty) \; .
\end{align}
\end{subequations}
The factor $\alpha \in \{-,+\}$ changes the sign in the integral equations depending on whether a bounded domain (interior problem) or a scatterer (exterior problem) is modelled. As mentioned above, these four representations will be used for testing only.

\section{Discretization} \label{sec.discretization}
The integral equations in \eqref{eq.collo_bie} and \eqref{eq:bie} have to be discretized in space and time. The temporal discretization will be performed with the CQM and the spatial discretization utilizes an IGA approach.

\subsection{Temporal discretization}\label{subsec.time_discretization}
In the present work, we refer to RK based convolution quadratures \cite{banjai2011runge}, which are an extension to the initial CQM from Lubich~\cite{lubich1988convolution}. For a general review of convolution quadrature and its applications we refer to \cite{lubich2004convolution}. Here, a variant is selected which essentially can be written as an usual transformation technique~\cite{schanz10a}.

The idea of convolution quadratures is the discretization of a convolution of two functions by a quadrature rule. Let us discretize the time interval $t \in [0,T]$ in $\nsteps$ equidistant time steps $\timestep$ such that $T=\nsteps \timestep$. Then, the quadrature formula is 
\begin{equation}\label{eq.cqm1}
  \cqmsol(\timevar) = (\timekernel * \timefunction)(\timevar)\rightsquigarrow\cqmsol(\timestepidx\timestep) \approx \cqmsol_\timestepidx = \transpose{\butcherb}\inverse{\butchera}\sum\limits_{\timestepidxidx=0}^{\timestepidx}\cqmWeights_{\timestepidx-\timestepidxidx}^{\timestep}(\frequencykernel) \myvec{\timefunction}_\timestepidxidx\qquad\timestepidx=0,\hdots,\nsteps\; .
\end{equation}
The matrix-valued quadrature weights $\cqmWeights_{\timestepidx-\timestepidxidx}^{\timestep}$ dependent on the Laplace transformed function $\frequencykernel=\laplacemap\{\timekernel\}$ and vector-valued quadrature points $\myvec{\timefunction}_\timestepidxidx=\timefunction(\timestepidxidx{\timestep}+\butcherc_{\rkstageidx}\timestep)$ with ${\rkstageidx}=1,\hdots,\rkstages$. The underlying $\rkstages-$stage RK method of (classical) order $\rkorder$ and stage order $\rkstageorder$ is given by its Butcher tableau
\begin{tabular}{c|c}
  $\butcherc$ & $\butchera$ \\ \hline
              & $\transpose{\butcherb}$
\end{tabular}
with $\butchera\in\Rn{\rkstages\times\rkstages}, \butcherb,\butcherc\in\Rn{\rkstages}$ and its stability function
\begin{equation}
  \stabfun(\zvar) = 1 + \zvar\transpose{\butcherb} \inverse{(\eye-\zvar\butchera)}\onevec\; ,
\end{equation}
where the abbreviation $\onevec := \transpose{[1,1,...,1]}$ is used for the vector of size $m$. We set the following requirements on the RK method
\begin{enumerate}
\item for all $\zvar\in\C$ s.t. $\real(\zvar)\leq 0$, it holds that $\abs{\stabfun(\zvar)}\leq 1$ and $(\eye-\zvar\butchera)$ is invertible,
\item for all $y\neq 0$, it holds that $\stabfun(iy)<1$, $\stabfun(\infty)=0$ and $\butchera$ is invertible.
\end{enumerate}
The latter entails that $\transpose{\butcherb}\inverse{\butchera}=[0,0,\hdots,1]$ holds. These conditions ensure A- and L-stability of the RK method.

The quadrature weights are defined by a closed, complex contour integral within the complex plane and can be approximated by a trapezoidal rule over $\ell=0,\hdots,\nfreqsteps-1$ complex frequency points 
\begin{equation}\label{eq.cqweights}
  \cqmWeights_{\timestepidx-\timestepidxidx}^{\timestep}(\frequencykernel)=\frac{1}{2\pi i}\oint\limits_{\abs{\zvar}=\cqmconst}\frequencykernel\left(\frac{\Delta(\zvar)}{\timestep}\right)\zvar^{\timestepidxidx-\timestepidx-1}d\zvar{\approx\frac{\cqmconst^{\timestepidxidx-\timestepidx}}{\nfreqsteps}\sum\limits_{\ell=0}^{\nfreqsteps-1}\frequencykernel\left( \frac{\Delta(\cqmconst \zeta^{\ell})}{\timestep}\right)\zeta^{(\timestepidxidx-\timestepidx)\ell}}\; ,
\end{equation}
with $\zeta=e^{2\pi i/\nfreqsteps}$, $\mathscr{R}\in(0,1)$ and the characteristic function $\Delta(\zvar)=\inverse{\left(\butchera+\frac{\zvar}{1-\zvar}\onevec\transpose{\butcherb}\right)}$. We substitute \eqref{eq.cqweights} into \eqref{eq.cqm1} and set the weights $\cqmWeights_{\timestepidx-\timestepidxidx}^{\timestep}(\frequencykernel)=0$ for negative indices $\timestepidx-\timestepidxidx<0$, which represents causality. By exchanging the order of summation, we write the discrete convolution with a weighted discrete Fourier transform (DFT) and its inverse transformation as
\begin{equation}
  (\timekernel*\timefunction)(\timestepidx\timestep) \approx\cqmsol_\timestepidx= \transpose{\butcherb}\inverse{\butchera}\underbrace{\frac{\cqmconst^{-\timestepidx}}{\nfreqsteps}\sum\limits_{\ell=0}^{\nfreqsteps-1}\Big[{\frequencykernel}\left( \frac{\Delta(\cqmconst \zeta^{\ell})}{\timestep}\right)\underbrace{\sum\limits_{\timestepidxidx=0}^{\nsteps-1}\cqmconst^{\timestepidxidx}\myvec{\timefunction}_\timestepidxidx\zeta^{\timestepidxidx\ell}}_{\text{weighted DFT} }\Big]\zeta^{-\timestepidx\ell}}_{\text{weighted inverse DFT } }.
\end{equation}
This formula can be applied respectively to \eqref{eq.collo_bie} and \eqref{eq:bie} to obtain semi-discrete boundary integral equations, i.e.,~the function $\freqkernel$ is then one of the semi-discrete layer potentials in the Laplace domain. However, the argument of the function is a matrix. Assuming, that the differential symbol $\frac{\Delta(\cqmconst\zeta^{\ell})}{\timestep}\in\C^{\rkstages\times\rkstages}$ has a full set of eigenvectors, we apply an eigenvalue decomposition $\frac{\Delta(\cqmconst\zeta^{\ell})}{\timestep}= \eigenvec\eigenval\inverse{\eigenvec}$ with $\eigenval=\diag{(\laplacevar_{0},\hdots,\laplacevar_{\rkstages-1})}$ and rewrite the composition of functions as
\begin{equation}
  {\frequencykernel}\left(\frac{\Delta(\cqmconst\zeta^{\ell})}{\timestep}\right)=\eigenvec\ \diag({\frequencykernel}(\laplacevar_{0}),\hdots,{\frequencykernel}(\laplacevar_{\rkstages-1}))\ \inverse{\eigenvec}.
\end{equation}
Applying the CQM in the above sketched form to \eqref{eq.collo_bie} and \eqref{eq:bie} results in the requirement to solve $\nfreqsteps\rkstages/2$ Laplace domain problems defined for $\rkstageidx=0,\hdots,\rkstages-1$ stages by
\begin{equation}\label{eq.elliptic1} 
  \begin{bmatrix}
    \freq{\SL} & -\freq{\DL}
  \end{bmatrix}
  \begin{bmatrix}
    \stress^*\\
    \displacement^*
  \end{bmatrix}  
  (\yb-\xb,\laplacevar_{\rkstageidx})
  =
  \begin{bmatrix}
    (\discretejump + \freq{\DL}) & -\freq{\SL}
  \end{bmatrix}
  \begin{bmatrix}
    \diridata^*\\
    \neumdata^*
  \end{bmatrix}
  (\yb-\xb,\laplacevar_{\rkstageidx})
\end{equation}
for collocation, and
\begin{equation}
  \begin{bmatrix}\label{eq.elliptic} 
    \freq{\SL} & -\freq{\DL}\\
    \hat{\myvec{K}}^{\prime} & \freq{\HSO}
  \end{bmatrix}
  (\yb-\xb,\laplacevar_{\rkstageidx})
  \begin{bmatrix}
    \stress^*  \\ \displacement^*
  \end{bmatrix}
  =
  \begin{bmatrix}
    (\frac{1}{2}\eye + \freq{\DL}) & -\freq{\SL} \\
    -\freq{\HSO}& (\frac{1}{2}\eye-\hat{\myvec{K}}^{\prime}) 
  \end{bmatrix}
  (\yb-\xb,\laplacevar_{\rkstageidx})
  \begin{bmatrix}
    \diridata^* \\ \neumdata^*
  \end{bmatrix}
\end{equation}
for the symmetric Galerkin formulation, respectively. In \eqref{eq.elliptic1} and \eqref{eq.elliptic}, $\freq{\SL}$, $\freq{\DL}$, $\hat{\myvec{K}}^{\prime}$, $\freq{\HSO}$ are the discrete integral operators in the Laplace domain and $\discretejump$ is the discrete integral free term. The boundary data $\diridata^*$ and $\neumdata^*$ at a stage $\timestepidx\timestep+\butcherc_{\rkstageidx}\timestep$ is constructed by the forward transformation of the coefficients obtained by a $\Ltwo{\boundary}$ projection of $\diribc(\xb,\timestepidx\timestep+\butcherc_{\rkstageidx}\timestep)$ and $\neumbc(\xb,\timestepidx\timestep+\butcherc_{\rkstageidx}\timestep)$, respectively.  Note that if $N=L$, an FFT-type algorithm can be used. The whole semi-discrete algorithm is given in~\Cref{algo:cqm}.
\begin{algorithm}[H]
  \caption{\label{algo:cqm}Pseudo code for the realization of the CQM for \eqref{eq.collo_bie} and \eqref{eq:bie} using the FFT}
  \begin{algorithmic}
    \For{$\compfreqidx = 0,\hdots,\nfreqsteps/2-1$} \Comment{complex frequency steps}
    \State
    $\frac{\Delta(\cqmconst\zeta^{\compfreqidx})}{\timestep}= \eigenvec\pmb{\Lambda}\inverse{\eigenvec} \qquad \text{where} \qquad\pmb{\Lambda}=\diag{(\laplacevar_{0},\hdots,\laplacevar_{\rkstages-1})}$
    \State
       $ \{\diridata^{**}\}_\rkstageidx = \sum\limits_{\timestepidxidx=0}^{\nsteps-1}\cqmconst^\timestepidxidx \{\diridata\}_{\timestepidxidx \rkstageidx}\zeta^{\timestepidxidx\compfreqidx}  \qquad \{\neumdata^{**}\}_\rkstageidx = \sum\limits_{\timestepidxidx=0}^{\nsteps-1}\cqmconst^\timestepidxidx \{\neumdata\}_{\timestepidxidx \rkstageidx}\zeta^{\timestepidxidx\compfreqidx}$
    \For{$\compfreqidxidx=0,\hdots,\rkstages-1$} \Comment{over the stages}
    \State  
        $\diridata^* =\sum\limits_{\rkstageidx=0}^{\rkstages-1}\inverse{\eigenvec}_{\compfreqidxidx\rkstageidx} \{\diridata^{**}\}_\rkstageidx  \qquad
        \neumdata^* =\sum\limits_{\rkstageidx=0}^{\rkstages-1}\inverse{\eigenvec}_{\compfreqidxidx\rkstageidx}\{\neumdata^{**}\}_\rkstageidx $
    \State 
        solve \eqref{eq.elliptic1} or \eqref{eq.elliptic} for $\myvec{\pressure}^*$ and $\myvec{\flux}^*$
        \State
      $  \{\myvec{{\pressure}^{**}}\}_{\compfreqidx \compfreqidxidx} =\sum\limits_{\rkstageidx=0}^{\rkstages-1}{\eigenvec}_{\compfreqidxidx\rkstageidx} \myvec{\pressure}^*\qquad\{\myvec{\flux}^{**}\}_{\compfreqidx \compfreqidxidx} =\sum\limits_{\rkstageidx=0}^{\rkstages-1}{\eigenvec}_{\compfreqidxidx\rkstageidx} \myvec{\flux}^*$
    \EndFor
    \EndFor
    \For{$\timestepidx=0,\hdots,\nsteps-1$}   \Comment{over time steps}
    \For{$\rkstageidx=0,\hdots,\rkstages-1$} \Comment{over the stages}
    \State ${\myvec{\pressure}}(\timestepidx\timestep+\butcherc_\rkstageidx\timestep) = \frac{2\cqmconst^{-\timestepidx}}{\nfreqsteps}\sum\limits_{\compfreqidx=0}^{\nfreqsteps/2-1}\{\myvec{{\pressure}^{**}}\}_{\compfreqidx \rkstageidx}\zeta^{-\timestepidx\compfreqidx}   
    \qquad 
    {\stress}(\timestepidx\timestep+\butcherc_\rkstageidx\timestep) = \frac{2\cqmconst^{-\timestepidx}}{\nfreqsteps}\sum\limits_{\compfreqidx=0}^{\nfreqsteps/2-1}\{\myvec{\flux}^{**}\}_{\compfreqidx \rkstageidx}\zeta^{-\timestepidx\compfreqidx}$
    \EndFor
    \EndFor
  \end{algorithmic}
\end{algorithm}

\subsection{Spatial discretization}\label{subsec.space_discretization}
Let the boundary be described exactly by an assembly of $\npatches$ non-overlapping patches $\boundary=\bigcup_{0\leq\patchidx<\npatches}\patch$. We assume the boundary to be Lipschitz continuous and closed. Each patch is equipped with a smooth, nonsingular and bijective diffeomorphism $\patchmap:\Rn{2}\mapsto\Rn{\mydimension}$
\begin{equation}\label{eq.geomap1}
  \patchmap:\reference\mapsto\patch
\end{equation}
that maps from a reference domain $\reference:=[0,1]^{2}$ to the physical domain $\patch$. 
In addition, we require that for any interface $\partial\boundary_{\patchidx_1}\bigcap\partial\boundary_{\patchidx_2}\neq\emptyset$ the mappings $\geomap_{\patchidx_1},\,\geomap_{\patchidx_2}$ coincide up to orientation. For each parametric direction in the reference domain $\midxidx=1,2$, let us introduce the vector $\uniqueknotvec_{\midxidx}=[\uniqueknot_{\midxidx,\,0}=0,\,\hdots,\,\uniqueknot_{\midxidx,\,\nelemsx_{\midxidx}}=1]$ of unique knot values, also called breakpoints. This nondecreasing sequence of unique parametric values divides each parametric direction $\midxidx$ into the intervals $\knotspan_{\midxidx,\,\elemidx_{\midxidx}}=[\uniqueknot_{\midxidx,\,\uniqueknotidx_{\midxidx}},\,\uniqueknot_{\midxidx,\,\uniqueknotidx_{\midxidx}+1}]$ of length $\elsize_{\midxidx,\,\elemidx_{\midxidx}}=\uniqueknot_{\midxidx,\,\uniqueknotidx_{\midxidx}+1}-\uniqueknot_{\midxidx,\,\uniqueknotidx_{\midxidx}}$ for $\elemidx_{\midxidx}=\uniqueknotidx_{\midxidx}=0,\,\hdots,\,\nelemsx_{\midxidx}-1$. Together, they form a cartesian grid in the reference domain as
\begin{equation}
  \mesh:=\{\elem_{\elemidx}=\knotspan_{1,\,\elemidx_{1}}\times\knotspan_{2,\,\elemidx_{2}}\vert\knotspan_{\midxidx,\,\elemidx_{\midxidx}}=[\uniqueknot_{\midxidx,\,\uniqueknotidx_{\midxidx}},\,\uniqueknot_{\midxidx,\,\uniqueknotidx_{\midxidx+1}}]\quad\text{for}\quad 0\leq\elemidx_{\midxidx},\uniqueknotidx_{\midxidx}\leq\nelemsx_{\midxidx}-1\}\; ,
\end{equation}
where $\elemidx$ indicates the global element index $\elemidx=\patchidx\nelemsx_1\nelemsx_2+\elemidx_2\nelemsx_1+\elemidx_1$. We call the image of $\mesh$ under $\patchmap$ a mesh of a patch $\patch$. Let us specify a set of $\nctrlpts_{\midxidx}$ polynomials of degree $\mydegree_{\midxidx}<\nctrlpts_{\midxidx}$ and continuity $C^{\delta_{\uniqueknotidx_{\midxidx}}}$ at the breakpoints by introducing the knot vector of a patch as
\begin{equation}\label{eq.knotvec_1}
  \knotvec_{\midxidx} = [ \underbrace{\uniqueknot_{\midxidx,0},\,\hdots,\,\uniqueknot_{\midxidx,0}}_{\multiplicity_{0}\ \text{times}},\,\underbrace{\uniqueknot_{\midxidx,1},\,\hdots,\,\uniqueknot_{\midxidx,1}}_{\multiplicity_{1}\ \text{times}},\,\hdots,\,\underbrace{\uniqueknot_{\midxidx,\nelemsx_{\midxidx}},\,\hdots,\,\uniqueknot_{\midxidx,\nelemsx_{\midxidx}}}_{\multiplicity_{\nelemsx_{\midxidx}}\ \text{times}}]\qquad\text{with}\qquad \sum_{\uniqueknotidx_{\midxidx}=0}^{\nelemsx_{\midxidx}}\multiplicity_{\uniqueknotidx_{\midxidx}}=\nctrlpts_{\midxidx}+\mydegree_{\midxidx}+1\; ,
\end{equation}
where $\multiplicity_{\uniqueknotidx_{\midxidx}}$ refers to the multiplicity of a knot value $\uniqueknot_{\midxidx,\uniqueknotidx_{\midxidx}}$, which determines $C^{\delta_{\uniqueknotidx_{\midxidx}}}$ by $\delta_{\uniqueknotidx_{\midxidx}}=\mydegree_{\midxidx}-\multiplicity_{\uniqueknotidx_{\midxidx}}$ for $0\leq \uniqueknotidx_{\midxidx}\leq\nelemsx_{\midxidx}$. 
Further, we use open knot vectors where $\multiplicity_{0}=\multiplicity_{\nelemsx_{\midxidx}}=\mydegree_{\midxidx}+1$. 
We relabel $\knotvec_{\midxidx}$ in terms of knots 
\begin{equation}
  \knotvec_{\midxidx} = [ \knot_{\midxidx,0} = ... = \knot_{\midxidx,\mydegree_{\midxidx}}=0 < \knot_{\midxidx,\mydegree_{\midxidx}+1}\leq ... \leq \knot_{\midxidx,\nctrlpts_{\midxidx}-1}< \knot_{\midxidx,\nctrlpts_{\midxidx}} = ... = \knot_{\midxidx,\nctrlpts_{\midxidx}+\mydegree_{\midxidx}}=1]\; .
\end{equation}
In the standard IGA framework, B-spline basis functions are constructed on the knot vector $\knotvec_{\midxidx}$ and used to define the geometric mapping $\patchmap$ by a linear combination with the tensor product grid of control points $\ctrlpoint_{\midx_1 \midx_2} = \transpose{[x_{1},x_{2},x_{\mydimension}]_{\midx_1 \midx_2}} \in \Rn{\mydimension}$. To parameterize conic sections exactly, we generalize \eqref{eq.geomap1} to NURBS mappings by a grid of associated weights $\ctrlweight_{\midx_1 \midx_2} \in \R$. 
For more details we refer to \cite{piegl2012nurbs}.

However, the resulting NURBS basis is generally non-univariate in the parametric directions, and supports span multiple elements. This variability complicates its integration into an existing BEM code based on a conventional polygonal boundary mesh. Therefore, we apply a series of basis transformations.

\paragraph{Local basis on the elements $\elem_{\elemidx}$}
To obtain a basis of uniform polynomial degree, we apply a degree elevation algorithm \cite{piegl2012nurbs} patch-wise. Let the global maximum geometric mapping degree be
\begin{equation}
  \mappingdegree =  \max_{0\leq\patchidx<\npatches}\max_{1 \leq \midxidx \leq 2} \mydegree_{\patchidx, \midxidx}\; .
\end{equation}
We iteratively insert the knots ${\uniqueknot}_{\midxidx,\,\insertionidx_{\midxidx}}$ where $\insertionidx_{\midxidx}= 0,\,\hdots,\,\nelemsx_{\midxidx} $ into $\knotvec_{\midxidx}$ until the first and last knot have a multiplicity of $\mappingdegree+1$. After every inserted knot, we update the control points $\ctrlpoint_{\midx_1 \midx_2} $ and the control weights $\ctrlweight_{\midx_1 \midx_2}$. 
For example, we elevate the control points in the first parametric direction by
\begin{equation}
  \elevated{\ctrlpoint}_{\insertionidx_1 {\midx}_2} = (1-\elevated{\insertioncoeff}_{\insertionidx_1})\ctrlpoint_{\insertionidx_1 \midx_2}^{} + \elevated{\insertioncoeff}_{\insertionidx_1}\ctrlpoint_{\insertionidx_1-1\, \midx_2}\qquad\text{with}\qquad \elevated{\insertioncoeff}_{\insertionidx_1} = \frac{\insertionidx_1}{1+\mappingdegree}\; .
\end{equation}
With $\Delta\mydegree_{g,1}=\mappingdegree-\mydegree_{g,1}$, we require $(\nelemsx_{1}+1)\Delta\mydegree_{g,1}$ insertions and the degree elevated control variables are defined on a grid of size $\elevated{\nctrlpts}_{1}\times\elevated{\nctrlpts}_{2}$ where $\elevated{\nctrlpts}_{\midxidx}=(\nelemsx_{\midxidx}+1)\Delta\mydegree_{g,\midxidx}+\sum\nolimits_{\insertionidx_{\midxidx}=0}^{\nelemsx_{\midxidx}}\multiplicity_{\insertionidx_{\midxidx}}$.
We now localize the elevated basis on the elements $\elem_{\elemidx}$ by the Bézier extraction \cite{borden2011isogeometric}. This is done by inserting the knots ${\uniqueknot}_{{\midxidx},\,\insertionidx_{\midxidx}}$ where $\insertionidx_{\midxidx}=1,\hdots,\nelemsx_{\midxidx}-1$ until every knot has the multiplicity of $\mappingdegree+1$. Thus, the degree elevated control variables $\elevated{\ctrlpoint}_{\elevated{\midx}_1 \elevated{\midx}_2}$ and $\elevated{\ctrlweight}_{\elevated{\midx}_1 \elevated{\midx}_2}$ are extracted by a knot insertion algorithm and reads, e.g.,~applied at the control points in the first parametric direction as
\begin{equation}\label{eq.insertioncoeffs}
  \extracted{\ctrlpoint}_{\insertionidx_{1} \elevated{\midx}_2} = (1-\extracted{\insertioncoeff}_{\insertionidx_{1}})\elevated{\ctrlpoint}_{\insertionidx_{1}-1\, \elevated{\midx}_2} + \extracted{\insertioncoeff}_{\insertionidx_{1}}\elevated{\ctrlpoint}_{\insertionidx_{1} \elevated{\midx}_2}\qquad\text{with}\qquad\extracted{\insertioncoeff}_{\insertionidx_{1}} = \begin{cases}
    1,&\insertionidx_{1}< \elevated{\nctrlpts}_{1}-\mappingdegree\; ,\\
    \frac{{\uniqueknot}_{1,\,\insertionidx_{1}}-\knot_{{1},\,\insertionidx_{1}}}{\knot_{1,\,\insertionidx_{1}+\mappingdegree}-\knot_{1,\,\insertionidx_{1}}},&\elevated{\nctrlpts}_{1}-\mappingdegree\leq \insertionidx_{1}< \elevated{\nctrlpts}_{1}\; ,\\
    0,&\insertionidx_{1}\geq \elevated{\nctrlpts}_{1}\; .
  \end{cases}
\end{equation}
After $\ninsertions_{1}=\sum\nolimits_{\uniqueknotidx_{1}=1}^{\nelemsx_{1}-1}(\mappingdegree+1)-(\multiplicity_{\uniqueknotidx_{1}}+\Delta\mydegree_{g,1})$ insertions the grid updates to $\extracted{\ctrlpoint}_{\extracted{\midx}_1 \elevated{\midx}_2}$ and $\extracted{\ctrlweight}_{\extracted{\midx}_1 \elevated{\midx}_2}$ with the index $0\leq\extracted{\midx}_{1}<\extracted{\nctrlpts}_{1}=\nelemsx_{1}(\mappingdegree+1)$. 
Applying these transformations to the two respective parametric directions and scaling the reference element to $\reference$ the basis functions can then be defined element-wise by Bernstein polynomials
\begin{align}
  \mappingbasis_{i,\mappingdegree}(\x_{\midxidx})=\frac{\mappingdegree !}{i !(\mappingdegree-i)!}\x_{\midxidx}^i(1-\x_{\midxidx})^{\mappingdegree - i} && \text{for} && i=0,\hdots,\mappingdegree && \text{and} && \x_{\midxidx} \in[0,1]
\end{align}
and the element-wise mapping reads $\elementmap:\extracted{\reference\to\boundary}_{\elemidx}$
\begin{equation}\label{eq.geomap2}
  \elementmap(\x_1,\x_2) = \frac{\sum\limits_{\extracted{\midx}_1=0}^{\mappingdegree}\sum\limits_{\extracted{\midx}_2=0}^{\mappingdegree}\mappingbasis_{\extracted{\midx}_1,\,\mappingdegree}(\x_1)\mappingbasis_{\extracted{\midx}_2,\,\mappingdegree}(\x_2)\extracted{\ctrlweight}_{\extracted{\midx}_1 \extracted{\midx}_2}\extracted{\ctrlpoint}_{\extracted{\midx}_1 \extracted{\midx}_2} }{\sum\limits_{\extracted{\midx}_1=0}^{\mappingdegree}\sum\limits_{\extracted{\midx}_2=0}^{\mappingdegree}\mappingbasis_{\extracted{\midx}_1,\,\mydegree_g}(\x_1)\mappingbasis_{\extracted{\midx}_2,\,\mydegree_g}(\x_2)\extracted{\ctrlweight}_{\extracted{\midx}_1 \extracted{\midx}_2}}=\frac{\myvec{A}(\x_1,\x_2)}{\ctrlweightfun(\x_1,\x_2)}\; .
\end{equation}
The partial derivative of \eqref{eq.geomap2} with respect to a reference coordinate $\x_i$, denoted as  
\begin{equation}
  \tvec^{(i)}(\x_1,\x_2) = \frac{\partial \extracted{\geomap}_\elemidx}{\partial x_{i}}  =  \frac{\myvec{A}^{(i)}(\x_1,\x_2) - \ctrlweightfun^{(i)}(\x_1,\x_2) \elementmap(\x_1,\x_2)}{\ctrlweightfun(\x_1,\x_2)}\;  ,
\end{equation}
implies the quotient rule \cite{piegl2012nurbs}. Due to the recursive nature of B-splines, they can be calculated iteratively by reducing the spline degree. 
For example, a partial derivative with respect to the $\x_1$ direction is then a mapping of order $\myvec{\mydegree}^{(1)} = [\mappingdegree-1, \mappingdegree]$ defined by $\tvec^{(1)}(\x_1,\x_2)$
with
\begin{subequations}
\begin{align}
  \myvec{A}^{(1)}(\x_1,\x_2) &=  \sum\limits_{\extracted{\midx}_1=0}^{\mappingdegree-1}\sum\limits_{\extracted{\midx}_2=0}^{\mappingdegree} \mappingbasis_{\extracted{\midx}_{1}, \mappingdegree-1}(\x_1)\mappingbasis_{\extracted{\midx}_2, \mappingdegree}(\x_2)\Delta\ctrlweight_{\extracted{\midx}_1 \extracted{\midx}_2}^{(1)}\Delta\ctrlpoint_{\extracted{\midx}_1 \extracted{\midx}_2}^{(1)}\; , \\
  \ctrlweightfun^{(1)}(\x_1,\x_2) &= \sum\limits_{\extracted{\midx}_1=0}^{\mappingdegree-1}\sum\limits_{\extracted{\midx}_2=0}^{\mappingdegree} \mappingbasis_{\extracted{\midx}_1, \mappingdegree-1}(\x_1)\mappingbasis_{\extracted{\midx}_2, \mappingdegree}(\x_2)\Delta\ctrlweight_{\extracted{\midx}_1\extracted{\midx}_2}^{(1)}\; ,
\end{align}
\end{subequations}
and the forward differences 
\begin{equation}
  \Delta\ctrlpoint_{\extracted{\midx}_1\,\extracted{\midx}_2}^{(1)} = (\ctrlpoint_{\extracted{\midx}_1+1\,\extracted{\midx}_2} - \ctrlpoint_{\extracted{\midx}_1\,\extracted{\midx}_2}) \mappingdegree \quad \text{and} \quad 
  \Delta\ctrlweight_{\extracted{\midx}_1\,\extracted{\midx}_2}^{(1)} = (\ctrlweight_{\extracted{\midx}_1+1\,\extracted{\midx}_2} - \ctrlweight_{\extracted{\midx}_1\,\extracted{\midx}_2}) \mappingdegree\; .
\end{equation}  
The surface measure $\surfmeasure_\elemidx(\x_1,\x_2):=\left\|\tvec^{(1)}(\x_1,\x_2) \times \tvec^{(2)}(\x_1,\x_2)\right\|$ enables the pull back of the surface integral on $\extracted{\boundary}_\elemidx$ to the reference domain $\reference$ and normalizes the outward normal vector 
\begin{equation}
  \nvec{\x_1,\x_2} = \frac{\tvec^{(1)}(\x_1,\x_2) \times \tvec^{(2)}(\x_1,\x_2)}{\surfmeasure_\elemidx(\x_1,\x_2)}\; .
\end{equation}
\paragraph{Solution basis for $\approximated{\pressure}(\xb,\timestepidx\timestep)$ and $\flux_h(\xb,\timestepidx\timestep)$}
For the B-spline bases of arbitrary degree $\mydegree$ and $\mydegree-1$ as the respective bases for the approximative solution $\approximated{\pressure}(\xb)$ and $\flux_h(\xb)$ let us consider the open knot vectors $\approxknotvec_{\midxidx}$ and $\approxknotvec_{\midxidx}'$ as in \eqref{eq.knotvec_1} with $1\leq\multiplicity_{\uniqueknotidx_{\midxidx}},\multiplicity_{\uniqueknotidx_{\midxidx}}'\leq\mydegree$ for $0< \uniqueknotidx_{\midxidx}<\nelemsx_{\midxidx}$. The sets of $\ncoeffs_{\midxidx}=-1-\mydegree+\sum\nolimits_{\uniqueknotidx_{\midxidx}=0}^{\nelemsx_{\midxidx}}\multiplicity_{\uniqueknotidx_{\midxidx}}$ and $\ncoeffs_{\midxidx}'=-\mydegree+\sum\nolimits_{\uniqueknotidx_{\midxidx}=0}^{\nelemsx_{\midxidx}}\multiplicity_{\uniqueknotidx_{\midxidx}}'$ degrees of freedom
\begin{equation}\label{eq.ansatzbasis}
  \{\refdom{\approxbasis}_{\midxb_{\midxidx},\mydegree}(\x_{\midxidx})\}_{\midxb_{\midxidx}=0}^{\ncoeffs_{\midxidx}-1}\qquad \text{and}\qquad \{\refdom{\approxbasis}_{\midxb_{\midxidx}',\mydegree-1}(\x_{\midxidx})\}_{\midxb_{\midxidx}'=0}^{\ncoeffs_{\midxidx}'-1}
\end{equation}
hold the respective approximating solutions at a point $\xb=\extracted{\geomap}_\elemidx(\x_1,\x_2)$ and a time step $\timestepidx\timestep$ 
\begin{subequations}
\begin{align}
    \approximated{\pressure}(\xb,\timestepidx\timestep)&=\sum\limits_{\midxb_{1}=0}^{\ncoeffs_{1}-1}\sum\limits_{\midxb_{2}=0}^{\ncoeffs_{2}-1}\diriunknown_{\midxb_{1}\midxb_{2}}({\timestepidx\timestep})\refdom{\approxbasis}_{{\midxb}_1,\mydegree}(\x_1)\refdom{\approxbasis}_{{\midxb}_2,\mydegree}(\x_2)\; , \\
    \flux_h(\xb,\timestepidx\timestep)&=\sum\limits_{\midxb_{1}'=0}^{\ncoeffs_{1}'-1}\sum\limits_{\midxb_{2}'=0}^{\ncoeffs_{2}'-1}\neumunknown_{\midxb_{1}'\midxb_{2}'}({\timestepidx\timestep})\refdom{\approxbasis}_{{\midxb}_{1}',\mydegree-1}(\x_1)\refdom{\approxbasis}_{{\midxb}_{2}',\mydegree-1}(\x_2)\; .
\end{align}
\end{subequations}

However, we also extract the Bézier elements from $\approxknotvec_{\midxidx}$ and $\approxknotvec_{\midxidx}'$ and write the Bézier extraction of the approximative bases in matrix form. To show the extraction from  $\approxknotvec_{\midxidx}$, we write the single knot insertion matrix
\begin{equation}
  \singleinsertionmat_{\midxidx,\insertionidx_{\midxidx}}=\begin{bmatrix}
    \extracted{\insertioncoeff}_1&1-\extracted{\insertioncoeff}_2&0&\cdots&&&0\\
    0&\extracted{\insertioncoeff}_2&1-\extracted{\insertioncoeff}_3&0&\cdots&&0\\
    \vdots&&&&&&\vdots\\
    0&\hdots&&&&\extracted{\insertioncoeff}_{(\ncoeffs_{\midxidx}+\insertionidx_{\midxidx}-1)}&1-\extracted{\insertioncoeff}_{(\ncoeffs_{\midxidx}+\insertionidx_{\midxidx})}\\
  \end{bmatrix}\qquad\in\Rn{(\ncoeffs_{\midxidx}+\insertionidx_{\midxidx}-1)\times(\ncoeffs_{\midxidx}+\insertionidx_{\midxidx})}
\end{equation}
and combine them for $\ninsertions_{\midxidx}$ insertions to $\singleinsertionmat_{\midxidx}=\singleinsertionmat_{\midxidx,0}\singleinsertionmat_{\midxidx,1}\cdots\singleinsertionmat_{\midxidx,\ninsertions_{\midxidx}-1}$. The same holds for $\approxknotvec_{\midxidx}'$ and $\ninsertions_{\midxidx}'$ insertions. The Bézier extraction of a patch then reads $\project_{\patchidx}  = \singleinsertionmat_{1}\times\singleinsertionmat_{2}$ and $\project'_{\patchidx}  = \singleinsertionmat'_{1}\times\singleinsertionmat'_{2}$ and allows us to respectively represent the $\ncoeffs_{\midxidx}$ and $\ncoeffs'_{\midxidx}$ coefficients on the B-spline basis by $\ncoeffs_{\midxidx}+\ninsertions_{\midxidx}$ and $\ncoeffs'_{\midxidx}+\ninsertions'_{\midxidx}$ coefficients on the Bernstein basis by 
\begin{equation}
 \diriunknown_{\patchidx}  = \project_{\patchidx}\extracted{\diriunknown}_{\patchidx}\qquad\text{and}\qquad\neumunknown_{\patchidx} = \project'_{\patchidx}\extracted{\neumunknown}_{\patchidx}\; .
\end{equation}
The Bernstein basis for approximating the solution is then uniformly defined on each element $\elem_{\elemidx}$ by
\begin{align*}
\extracted{\refdom{\approxbasis}}_{\localdofidx_{\midxidx},\mydegree}(\x_{\midxidx})=\frac{\mydegree !}{\localdofidx_{\midxidx} !(\mydegree-\localdofidx_{\midxidx})!}\x_{\midxidx}^{\localdofidx_{\midxidx}}(1-\x_{\midxidx})^{\mydegree - \localdofidx_{\midxidx}}&& \text{for} && \localdofidx_{\midxidx}=0,\hdots,\mydegree && \text{and} && \x_{\midxidx} \in[0,1]\; .
\end{align*}
\paragraph{Equation system in the Laplace domain} The open knot vectors of $\approxknotvec_{\midxidx}$ and $\approxknotvec'_{\midxidx}$ imply discontinuities between patches. For the approximation of the Dirichlet field, we consider a continuity of $\cont{0}{\boundary}$ between patches. Therefore, we introduce a global summation matrix $\glue=[g_{ij}]$ with $g_{ij}\in\{0,\,1\}$ that enforces the $\cont{0}{\boundary}$ continuity on the patch interfaces by the multiplication
\begin{equation}\label{eq.transf}
  \transf:=\transpose{\project}\glue\; ,
\end{equation}
where $\project=\diag(\project_{0},\hdots,\project_{\npatches-1})$ is the assembly of the patch-wise basis transformations.
The continuity of the Neumann field depends on the tangent space of the geometry and may be $\cont{-1}{\boundary}$. In that case $\glue'=\eye$ holds, i.e.,~$\transf' = \diag(\project'_{0},\hdots,\project'_{\npatches-1})$.
In the case of collocation, we interpolate the solution between the mapped points $\collopt=\extracted{\geomap}_\elemidx(\grevillept)$ where $\grevillept=\transpose{[\greville_{1,\midxb_{1}},\greville_{2,\midxb_{2}}]}$ and $\greville_{\midxidx,\midxb_{\midxidx}}$ denotes the $\midxb_{\midxidx}=0,\hdots,\nctrlpts_{\midxidx}$ Greville abscissae defined on $\knotvec_{\midxidx}$ \cite{greville1964numerical}. The equation system \eqref{eq.elliptic1} is then written as the transformed matrix equation
\begin{equation}
  \begin{bmatrix}
    \freq{\SL}\transf & -\freq{\DL}\transf
  \end{bmatrix}
  \begin{bmatrix}
    \myvec{\flux}^*\\
    \displacement^*
  \end{bmatrix}  
  (\yb-\collopt,\laplacevar)
  =
  \begin{bmatrix}
    (\discretejump + \freq{\DL})\transf & -\freq{\SL}\transf
  \end{bmatrix}
  \begin{bmatrix}
    \diridata^*\\
    \neumdata^*
  \end{bmatrix}
  (\yb-\collopt,\laplacevar)\; .
\end{equation}
To define the assembly of the operators, let us rewrite the basis functions of an element in tensor product form $\extracted{\refdom{\approxbasis}}_{\localdofidx,\mydegree}(\xr)=\extracted{\refdom{\approxbasis}}_{\localdofidx_1,\mydegree}(\x_{1})\extracted{\refdom{\approxbasis}}_{\localdofidx_2,\mydegree}(\x_{2})$ where $\localdofidx=\localdofidx_2(\mydegree+1)+\localdofidx_1$. Further, let us introduce the global index $\globaldofidx=\elemsupportidx(\mydegree+1)^2+\localdofidx$ where $\elemsupportidx=0,\hdots,\setdim{\setsupportelems}$ and $\setsupportelems(\globaldofidx)$ is the set of support elements of the $\globaldofidx\textsuperscript{th}$ basis function. 
A matrix entry of the Laplace domain operators in the discrete setting reads for a globally indexed collocation point $\collopt_{\globalcolloidx}$ as  
\begin{align*}
  {\freq{\SL}}[\globalcolloidx ,\globaldofidx' ] &=  \sum_{\extracted{\boundary}_{f} \in \setsupportelems(\globaldofidx')} \int\nolimits_{\extracted{\geomap}_f(\reference)} \freqfundsol\left(\yb-\collopt_{\globalcolloidx}, \laplacevar\right)\extracted{\approxbasis}_{\globaldofidx', \mydegree-1}(\yb)d\extracted{\boundary}_{f}\\
                                                 &= \sum_{\extracted{\boundary}_{f} \in \setsupportelems(\globaldofidx')}  \int\nolimits_{\reference}\freqfundsol\left(\extracted{\geomap}_{f}(\yr)-\extracted{\geomap}_\elemidx(\grevillept),\laplacevar\right)\extracted{\refdom{\approxbasis}}_{\localdofidx', \mydegree-1}(\yr)\surfmeasure_{f}(\yr) d{\yr}\; ,\\
  {\freq{\DL}}[\globalcolloidx', \globaldofidx] &= \sum_{\extracted{\boundary}_f \in \setsupportelems(\globaldofidx)}\int\nolimits_{\extracted{\geomap}_{f}(\reference)} \neumtrace{\yb}\left(\freqfundsol\left(\yb-\collopt_{\globalcolloidx'},\laplacevar\right)\right)\extracted{\approxbasis}_{\globaldofidx, \mydegree}(\yb)d\extracted{\boundary}_f\\
                                                &= \sum_{\extracted{\boundary}_f \in \setsupportelems(\globaldofidx)}\int\nolimits_{\reference} \neumtrace{\yb}\left(\freqfundsol\left(\extracted{\geomap}_{f}(\yr)-\extracted{\geomap}_\elemidx(\grevillept),\laplacevar\right)\right)\extracted{\refdom{\approxbasis}}_{\localdofidx, \mydegree}(\yr)\surfmeasure_{f}(\yr) d{\yr}\; .
\end{align*}
For the Galerkin formulation, the equation system \eqref{eq.elliptic} is then written as
\begin{equation}
  \begin{bmatrix}
    \transpose{(\transf')}{{\freq{\SL}}}\transf' & -  \transpose{(\transf')}{\freq{\DL}}\transf \\
    \transpose{\transf}{\freq{\DL}}'\transf'  &  \transpose{\transf}{\freq{\HSO}}\transf
  \end{bmatrix}(\yb-\xb,\laplacevar)                                                                             \begin{bmatrix}
    {\neumunknown^*}\\
    {\diriunknown^*}
  \end{bmatrix}
  =
  \begin{bmatrix}
    \transpose{(\transf')}({{\frac{1}{2}\eye}}+{\freq{\DL}})\transf & - \transpose{(\transf')}{\freq{\SL}}\transf \\
    - \transpose{\transf}{\freq{\HSO}}\transf &\transpose{\transf}(\frac{1}{2}\eye-{\freq{\DL}}')\transf' 
  \end{bmatrix}(\yb-\xb,\laplacevar)
  \begin{bmatrix}
    \diridata^*\\
    \neumdata^*
  \end{bmatrix} \; ,
\end{equation}
where the Laplace domain operators in the discrete setting read
\begin{align*}
  {\freq{\SL}}[\globaldofidx_x' ,\globaldofidx_y' ] &=  \sum_{\extracted{\boundary}_{\elemidx} \in S(\globaldofidx_x')} \sum_{\extracted{\boundary}_f \in S(\globaldofidx_y')}\int\nolimits_{\extracted{\geomap}_\elemidx(\reference)}\int\nolimits_{\extracted{\geomap}_{f}(\reference)} \freqfundsol\left(\yb-\xb, \laplacevar\right)\extracted{\approxbasis}_{\globaldofidx_x', \mydegree-1}(\xb)\extracted{\approxbasis}_{\globaldofidx_y', \mydegree-1}(\yb)d\extracted{\boundary}_fd\extracted{\boundary}_{\elemidx}\\
                                      &= \sum_{\extracted{\boundary}_{\elemidx} \in S(\globaldofidx_x')} \sum_{\extracted{\boundary}_f \in S(\globaldofidx_y')} \int\nolimits_{\reference}\int\nolimits_{\reference}\freqfundsol\left(\extracted{\geomap}_{f}(\yr)-\extracted{\geomap}_\elemidx(\xr),\laplacevar\right)\extracted{\refdom{\approxbasis}}_{\localdofidx_x', \mydegree-1}(\xr)\extracted{\refdom{\approxbasis}}_{\localdofidx_y', \mydegree-1}(\yr)\surfmeasure_\elemidx(\xr)\surfmeasure_{f}(\yr) d{\yr}d{\xr}\; ,\\
  {\freq{\DL}}[\globaldofidx', \globaldofidx] &= \sum_{\extracted{\boundary}_{\elemidx} \in S(\globaldofidx')} \sum_{\extracted{\boundary}_f \in S(\globaldofidx)}\int\nolimits_{\extracted{\geomap}_\elemidx(\reference)}\int\nolimits_{\extracted{\geomap}_{f}(\reference)} \neumtrace{\yb}\left(\freqfundsol\left(\yb-\xb,\laplacevar\right)\right)\extracted{\approxbasis}_{\globaldofidx', \mydegree-1}(\xb)\extracted{\approxbasis}_{\globaldofidx, \mydegree}(\yb)d\extracted{\boundary}_{f}d\extracted{\boundary}_{\elemidx}\\
                                      &= \sum_{\extracted{\boundary}_{\elemidx} \in S(\globaldofidx')} \sum_{\extracted{\boundary}_f \in S(\globaldofidx)}\int\nolimits_{\reference}\int\nolimits_{\reference} \neumtrace{\yb}\left(\freqfundsol\left(\extracted{\geomap}_{f}(\yr)-\extracted{\geomap}_\elemidx(\xr),\laplacevar\right)\right)\extracted{\refdom{\approxbasis}}_{\localdofidx', \mydegree-1}(\xr)\extracted{\refdom{\approxbasis}}_{\localdofidx, \mydegree}(\yr)\surfmeasure_\elemidx(\xr)\surfmeasure_{f}(\yr) d{\yr}d{\xr} \;  ,\\
  {\freq{\DL}}'[\globaldofidx, \globaldofidx']&= \sum_{\extracted{\boundary}_{\elemidx} \in S(\globaldofidx)} \sum_{\extracted{\boundary}_f \in S(\globaldofidx')}\int\nolimits_{\extracted{\geomap}_\elemidx(\reference)}\int\nolimits_{\extracted{\geomap}_{f}(\reference)} \neumtrace{\xb}\left(\freqfundsol\left(\yb-\xb,\laplacevar\right)\right)\extracted{\approxbasis}_{\globaldofidx, \mydegree}(\xb)\extracted{\approxbasis}_{\globaldofidx', \mydegree-1}(\yb)d\extracted{\boundary}_{f}d\extracted{\boundary}_{\elemidx}\\
                                      &= \sum_{\extracted{\boundary}_{\elemidx} \in S(\globaldofidx)} \sum_{\extracted{\boundary}_f \in S(\globaldofidx')}\int\nolimits_{\reference}\int\nolimits_{\reference} \neumtrace{\xb}\left(\freqfundsol\left(\extracted{\geomap}_{f}(\yr)-\extracted{\geomap}_\elemidx(\xr),\laplacevar\right)\right)\extracted{\refdom{\approxbasis}}_{\localdofidx, \mydegree}(\xr)\extracted{\refdom{\approxbasis}}_{\localdofidx', \mydegree-1}(\yr)\surfmeasure_\elemidx(\xr)\surfmeasure_{f}(\yr) d{\yr}d{\xr} \;  ,\\
  {\freq{\HSO}}[\globaldofidx_x, \globaldofidx_y] &=  -\sum_{\extracted{\boundary}_{\elemidx} \in S(\globaldofidx_x)} \sum_{\extracted{\boundary}_f \in S(\globaldofidx_y)}\int\nolimits_{\extracted{\geomap}_\elemidx(\reference)}\int\nolimits_{\extracted{\geomap}_{f}(\reference)}\neumtrace{\xb}\left(\neumtrace{\yb}\left( \freqfundsol\left(\yb-\xb,\laplacevar\right)\right)\right)\extracted{\approxbasis}_{\globaldofidx_x, \mydegree}(\xb)\extracted{\approxbasis}_{\globaldofidx_y, \mydegree}(\yb)d\extracted{\boundary}_{f}d\extracted{\boundary}_{\elemidx}\\
                                          &=  -\sum_{\extracted{\boundary}_{\elemidx} \in S(\globaldofidx_x)} \sum_{\extracted{\boundary}_f \in S(\globaldofidx_y)}\int\nolimits_{\reference}\int\nolimits_{\reference} \neumtrace{\xb}\left(\neumtrace{\yb}\left(\freqfundsol\left(\extracted{\geomap}_{f}(\yr)-\extracted{\geomap}_\elemidx(\xr),\laplacevar\right)\right)\right)\extracted{\refdom{\approxbasis}}_{\localdofidx_x, \mydegree}(\xr)\extracted{\refdom{\approxbasis}}_{\localdofidx_y, \mydegree}(\yr)\surfmeasure_\elemidx(\xr)\surfmeasure_{f}(\yr) d{\yr}d{\xr}\; .
\end{align*}
The fundamental solutions  for acoustics and elastodynamics can be found in textbooks, e.g.,~\cite{gaul03,steinbach2007numerical}.

For the forward transformations
\begin{align} 
  \{\diridata^{**}\}_{\rkstageidx} = \sum\limits_{\timestepidxidx=0}^{\nsteps-1}\cqmconst^\timestepidxidx \{\diridata\}_{\timestepidxidx j}\zeta^{\timestepidxidx\ell}  &&\text{and}&& \{\neumdata^{**}\}_{\rkstageidx} = \sum\limits_{\timestepidxidx=0}^{\nsteps-1}\cqmconst^\timestepidxidx \{\neumdata\}_{\timestepidxidx j}\zeta^{\timestepidxidx\ell}
\end{align}
      as in \Cref{algo:cqm}, we compute the coefficient sets $\{\myvec{g}_{\dirisub}\}_{\timestepidxidx j}$ and $\{\myvec{g}_{\neumsub}\}_{\timestepidxidx j}$ by a $\Ltwo{\boundary}$ projection of the given functions evaluated at the RK stages $\diribc(\timestepidx\timestep+\butcherc_{\rkstageidx}\timestep)$ and $\neumbc(\timestepidx\timestep+\butcherc_{\rkstageidx}\timestep)$  with $j=0,\hdots,\rkstages-1$, respectively by
\begin{align}\label{eq.l2projection}
  \transpose{\transf}{\mass}\transf \ \diridata= \transpose{\transf}{\rhs}&&\text{and}&&\transpose{(\transf')}{\mass}'\transf' \ \neumdata= \transpose{(\transf')}{\rhs}'
\end{align}
with the mass matrices
\begin{align*}
  {\mass}[\globaldofidx_x,\globaldofidx_y] &= \sum_{\extracted{\boundary}_{\elemidx} \in S(\globaldofidx_x)\cap\diriboundary}\ \sum_{\extracted{\boundary}_f \in S(\globaldofidx_y)\cap\diriboundary}  \int\nolimits_{\extracted{\geomap}_\elemidx(\reference)}\int\nolimits_{\extracted{\geomap}_{f}(\reference)} \extracted{\approxbasis}_{\globaldofidx_x, \mydegree}(\xb)\extracted{\approxbasis}_{\globaldofidx_y, \mydegree}(\yb)d\extracted{\boundary}_{f}d\extracted{\boundary}_{\elemidx}\\
                                     &= \sum_{\extracted{\boundary}_{\elemidx} \in S(\globaldofidx_x)\cap\diriboundary}\ \sum_{\extracted{\boundary}_f \in S(\globaldofidx_y)\cap\diriboundary} \int\nolimits_{\reference}\int\nolimits_{\reference} \extracted{\refdom{\approxbasis}}_{\localdofidx_x, \mydegree}(\xr)\extracted{\refdom{\approxbasis}}_{\localdofidx_y, \mydegree}(\yr)\surfmeasure_\elemidx(\xr)\surfmeasure_{f}(\yr) d{\yr}d{\xr}\; ,\\
  {\mass}' [\globaldofidx_x',\globaldofidx_y'] &= \sum_{\extracted{\boundary}_{\elemidx} \in S(\globaldofidx_x')\cap\neumboundary}\ \sum_{\extracted{\boundary}_f \in S(\globaldofidx_y')\cap\neumboundary}  \int\nolimits_{\extracted{\geomap}_\elemidx(\reference)}\int\nolimits_{\extracted{\geomap}_{f}(\reference)} \extracted{\approxbasis}_{\globaldofidx_x', \mydegree-1}(\xb)\extracted{\approxbasis}_{\globaldofidx_y', \mydegree-1}(\yb)d\extracted{\boundary}_{f}d\extracted{\boundary}_{\elemidx}\\
                             &= \sum_{\extracted{\boundary}_{\elemidx} \in S(\globaldofidx_x')\cap\neumboundary}\ \sum_{\extracted{\boundary}_f \in S(\globaldofidx_y')\cap\neumboundary} \int\nolimits_{\reference}\int\nolimits_{\reference} \extracted{\refdom{\approxbasis}}_{\localdofidx_x', \mydegree-1}(\xr)\extracted{\refdom{\approxbasis}}_{\localdofidx_y', \mydegree-1}(\yr)\surfmeasure_\elemidx(\xr)\surfmeasure_{f}(\yr) d{\yr}d{\xr}\; ,
\end{align*}
and the functionals
\begin{align*}
  {\rhs}[\globaldofidx] &=  \sum_{\extracted{\boundary}_{\elemidx} \in S(\globaldofidx)\cap\diriboundary}  \int\nolimits_{\extracted{\geomap}_\elemidx(\reference)}\extracted{\approxbasis}_{\globaldofidx, \mydegree}(\xb)\diribc(\xb,\timestepidx\timestep+\butcherc_{\rkstageidx}\timestep)d\extracted{\boundary}_{\elemidx}=  \int\nolimits_{\reference}\extracted{\approxbasis}_{\localdofidx, \mydegree}(\xr)\surfmeasure_\elemidx(\xr)\diribc\left(\extracted{\geomap}_\elemidx(\xr),\timestepidx\timestep+\butcherc_{\rkstageidx}\timestep\right)d{\xr}\; ,\\
  {\rhs}' [\globaldofidx'] &=  \sum_{\extracted{\boundary}_{\elemidx} \in S(\globaldofidx')\cap\neumboundary}  \int\nolimits_{\extracted{\geomap}_\elemidx(\reference)}\extracted{\approxbasis}_{\globaldofidx', \mydegree-1}(\xb)\neumbc(\xb,\timestepidx\timestep+\butcherc_{\rkstageidx}\timestep)d\extracted{\boundary}_{\elemidx}=  \int\nolimits_{\reference}\extracted{\refdom{\approxbasis}}_{\localdofidx', \mydegree-1}(\xr)\surfmeasure_\elemidx(\xr)\neumbc\left(\extracted{\geomap}_\elemidx(\xr),\timestepidx\timestep+\butcherc_{\rkstageidx}\timestep\right)d{\xr}\; .
\end{align*}
The weakly singular integrals are treated with the Duffy transformation \cite{duffy2004transform}. Standard Gau\ss{} integration is used to integrate regular integrals numerically and the equation system is solved directly via a LU-decomposition.

\section{Discussion on expected rate of convergence} \label{sec.optimal_convergence}
Below some convergence studies will be presented. As the problem is discretized in space and time both discretizations will introduce errors and, hence, a combined $L^2$-error in space and time will be utilized. A convergence rate measured in such a norm will then determine the behavior with respect to both discretizations. Hence, the overall rate will be governed by the slower rate of convergence either in space or in time. That is why in the following both behaviors will be discussed separately.

\subsection{Convergence rate in time}
The approximation quality of the CQM in terms of convergence rates has been proven for a family of RK convolution quadratures by the energy conservation of Fourier transforms (Parseval's identity) \cite{banjai2012runge}. The application to an indirect Dirichlet problem was treated in \cite{banjai2011runge} where convergence rates are primarily dependent on the behavior of the Laplace transformed integral kernels of the boundary integral operators, more specifically, on their inverse. The error analysis in \cite{banjai2011runge} gives an error estimate of the form:
For $f:X\to Y$
 \begin{align}\label{eq.energynormestimate}
    \funnorm{(f*\timefunction) - (\approximated{f}*\timefunction)}{Y}\leq C &\timestep^{\min\{\rkorder,\rkstageorder+1-\mu+\nu\}}\left(\funnorm{\frac{\partial^\ell}{\partial\timevar^\ell}\timefunction(\cdot,0)}{X} + \int\limits_{0}^{\timevar}\funnorm{\frac{\partial^{\ell+1}}{\partial\timevar^{\ell+1}}\timefunction(\cdot,\tau)}{X}d\tau\right).
  \end{align}
The estimate above is sharp for $\timefunction(\cdot,\timevar)\in\Ltwo{\R}\vert\timefunction(\cdot,\timevar)\equiv 0$ for $\timevar<0$ and vanishing derivatives $\timefunction(\cdot,0)=\frac{\partial}{\partial\timevar}\timefunction(\cdot,0)=\hdots=\frac{\partial^{(\ell-1)}}{\partial\timevar^{(\ell-1)}}\timefunction(\cdot,0)=0$ with $\ell>\max\{\rkorder+\mu+1,\mydegree_{\timevar},\rkstageorder+1\}$.  The parameters $\mu$ and $\nu$ determine the growth behavior of $\frequencykernel=\laplacemap\{\timekernel\}$ by 
\begin{equation}\label{eq.kernel1}
  \abs{\frequencykernel(\laplacevar)}\leq M(\sigma_0)\frac{\abs{\laplacevar}^{\mu}}{(\real(\laplacevar))^\nu}\qquad\forall\real(\laplacevar)>\sigma_0>0
\end{equation}
for some $\mu\in\R$, $\nu\geq 0$ and a positive bounding factor $M(\sigma_0)>0$.  In \cite[Lemma 7]{banjai2011runge} and \cite{banjai2022integral}, it was proven that
\begin{equation}\label{eq.cqm_estimates}
    \funnorm{\inverse{\freq{\slp}}(\laplacevar)}{}\leq C \frac{\abs{\laplacevar}^2}{\real(\laplacevar)}\max\{1,\abs{\laplacevar}^{-1}\}\quad\text{and}\quad\funnorm{\inverse{\freq{\hsop}}(\laplacevar)}{}\leq C \frac{\abs{\laplacevar}}{\real(\laplacevar)}\max\{1,\abs{\laplacevar}^{-2}\}
  \end{equation}
hold, which allows us to give the convergence rate if the given right hand side $g$ is smooth enough. Unfortunately, the values $\mu$ and $\nu$ cannot be determined, neither for the inverse of our combined operator in \eqref{eq.elliptic} nor for the inverses of the double layer operator and its adjoint. Considering the estimates \eqref{eq.cqm_estimates} and taking the inverse of the single layer operator as dominant, we expect for our problem the stage order $\rkstageorder$ as convergence rate. Note that we will construct the right hand side such that $\ell$ fulfils the above criterion for the proof of the inverse single layer operator.

\subsection{Convergence rate in space}
The overall convergence rate is restricted as well by the accuracy of the computed solutions in the Laplace domain. We assume an elliptic behavior of the single layer and the hyper singular operator and refer to \cite{steinbach2007numerical,sauter2011boundary} for the considered error estimates. Let us write the weak formulation of the problem in the Laplace domain as a sesquilinear form $a(\pressure,\flux;\testD,\testN)=F(\testD,\testN)$ with
\begin{subequations}
\begin{align}
    a(\pressure,\flux;\testD,\testN) & =\innerprod{\freq{\slp}\flux}{\testN}{\diriboundary}-\innerprod{\freq{\dlp}\pressure}{\testN}{\diriboundary}+\innerprod{\freq{\dlp}'\flux}{\testD}{\neumboundary}+\innerprod{\freq{\hsop}\pressure}{\testD}{\neumboundary}\; ,\\
    F(\testD,\testN) & = \innerprod{(\frac{1}{2}\idty+\freq{\dlp}){\timefunction}_{\dirisub}-\freq{\slp}{\timefunction}_{\neumsub}}{\testN}{\diriboundary}+\innerprod{(\frac{1}{2}\idty-\freq{\dlp}'){\timefunction}_{\neumsub}-\freq{\hsop}{\timefunction}_{\dirisub}}{\testD}{\neumboundary}\; ,
\end{align}
\end{subequations}
which is satisfied for all $(\testD,\testN)\in{\hilbert}^{1/2}(\neumboundary)\times{\hilbert}^{-1/2}(\diriboundary)$ and holds the unique solution $(\pressure,\flux)\in{\hilbert}^{1/2}(\neumboundary)\times{\hilbert}^{-1/2}(\diriboundary)$. An approximate solution $(\approximated{\pressure},\approximated{\flux})\in\gspace_{0}(\neumboundary)\cap{\hilbert}^{1/2}(\neumboundary)\times\gspace_{1}(\diriboundary)$ satisfies the Galerkin variational formulation
\begin{equation}
  a(\approximated{\pressure},\approximated{\flux};\approximated{\testD},\approximated{\testN})=F(\approximated{\testD},\approximated{\testN})
\end{equation}
for all test functions $(\approximated{\testD},\approximated{\testN})\in\gspace_{0}(\neumboundary)\cap{\hilbert}^{1/2}(\neumboundary)\times\gspace_{1}(\diriboundary)$. We assume that $\pressure$ satisfies $\hilbert^{\smoothness}(\domain)$ and $\hilbert^{\smoothness-1/2}(\boundary)$ for some $\smoothness>3/2$ and that $\flux$ is in a piece-wise defined Hilbert space $\hilbert^{\smoothness-3/2}_{pw}(\boundary)$. For sufficiently regular ${\timefunction}_{\dirisub}\in\hilbert^{\smoothness-1/2}(\boundary)$ and ${\timefunction}_{\neumsub}\in\hilbert^{\smoothness-3/2}(\boundary)$ and conforming finite function spaces $\gspace_{0}(\boundary)$ and $\gspace_{1}(\boundary)$, the error estimate of a mixed boundary element problem corresponds to those of the pure Dirichlet and the Neumann problems \cite{steinbach2007numerical} and read respectively as
\begin{subequations}
\begin{align}
  &\funnorm{\pressure-\approximated{\pressure}}{\hilbert^{\sigma}(\boundary)}\lesssim \elsize^{\smoothness-\sigma}\abs{\pressure}_{\hilbert^{\smoothness}(\boundary)}\quad&\text{for}\quad &1/2\leq\smoothness\leq\mydegree+1\quad &\text{and}\quad& -1\leq\sigma\leq 1/2\; ,\label{eq.diriestimate}\\
  &\funnorm{\flux-\approximated{\flux}}{\hilbert^{\sigma}(\boundary)}\lesssim\elsize^{\smoothness-\sigma}\abs{\flux}_{\hilbert_{pw}^{\smoothness}(\boundary)}\quad&\text{for}\quad &-1/2\leq\smoothness\leq\mydegree\quad &\text{and}\quad&-2\leq\sigma\leq-1/2\; .\label{eq.neumestimate}
\end{align}
\end{subequations}
The perturbation of the problem by the approximation of the given functions does not lower the rate of optimal convergence due to a similar convergence rate and the best approximation property of the $L^2$-projection. Hence, under the conditions given above, we can expect convergence rates in the $\Ltwo{\boundary}$ norm of at most $\mydegree + 1$ for the Dirichlet data and $\mydegree$ for the Neumann data, respectively.

Spline spaces have been analyzed by \cite{wolf2020analysis,dolz2018fast,da2014mathematical}, thus, for further details we refer to the cited literature. However, the analysis is based on projections $\projector^{0}_{\mydegree,\,\knotvec}:\Ltwo{0,1}\mapsto S_{\mydegree}(\knotvec)$ in the reference domain where $S_{\mydegree}(\knotvec)$ is a $\nctrlpts$-dimensional spline space of degree $\mydegree$ on $\knotvec$. The classical polynomial estimate \cite{brenner2008mathematical}
\begin{equation}\label{eq.estimate1}
  \funnorm{f_0-\projector^{0}_{\mydegree,\,\knotvec}f_0}{\Ltwo{\knotspan}}\lesssim h^\smoothness\funnorm{f_0}{\hilbert^\smoothness({\knotspan)}}\qquad\forall f_0\in\hilbert^{\smoothness}(0,1)\quad\text{and}\quad 1\leq\smoothness\leq\mydegree+1
\end{equation}
holds for such projections at the knot interval $\knotspan=[\uniqueknot_{i},\uniqueknot_{i+1}]$, $i=0,\hdots,\nctrlpts-1$. The projections with its generalized derivative is spline preserving and reads $\projector_{\mydegree,\,\knotvec}^{1}:\Ltwo{0,1}\to S_{\mydegree-1}(\knotvec')$ where $\knotvec'$ is referred to as the truncated knot vector. From there, one finds
\begin{equation}\label{eq.estimate2}
  \funnorm{f_1-\projector^{1}_{\mydegree,\,\knotvec}f_1}{\Ltwo{\knotspan}}\lesssim h^\smoothness\funnorm{f_1}{\hilbert^\smoothness({\knotspan)}}\qquad\forall f_1\in\hilbert^{\smoothness}(0,1)\quad\text{and}\quad 0\leq\smoothness\leq\mydegree\; .
\end{equation}
By tensor product arguments the estimate also applies to the reference patch. Due to the smoothness and the bijectivity of the mappings $\patchmap$ and enforcing the interpolation property at the patch interfaces, the estimate can be defined on the boundary by
\begin{equation}\label{eq.approx01}
  \norm{f_{\dirineum}-\projector_{\boundary}^{\dirineum}f_{\dirineum}}_{L^2(\boundary)} 
  \lesssim \elsize^{\smoothness} \norm{f_{\dirineum}}{\hilbert_{pw}^{\smoothness}(\boundary)}
  \quad \text{with} \quad
  \begin{cases}
    2 \leq \smoothness \leq \mydegree + 1 & \text{for } \dirineum = 0 \;, \\
    0 \leq \smoothness \leq \mydegree     & \text{for } \dirineum = 1 \;,
  \end{cases}
\end{equation}
for any $f_{\dirineum}\in\hilbert_{pw}^{\smoothness}(\boundary)$.
Thus, spline approximations in the global spline spaces
\begin{align*}
  \gspace_{0}(\boundary)&:=\{f_{0}:(f_{0}\vert_{\patch}\circ\funmap_{\patchidx})\in S_{\mydegree_{1}}(\knotvec_{1})\times S_{\mydegree_{2}}(\knotvec_{2})\quad\text{for}\quad 0\leq\patchidx<\npatches\}\; ,\\
  \gspace_{1}(\boundary)&:=\{f_{1}:\surfmeasure_{\patchidx}(f_{1}\vert_{\patch}\circ\funmap_{\patchidx})\in S_{\mydegree_{1}-1}(\knotvec_{1}')\times S_{\mydegree_{2}-1}(\knotvec_{2}')\quad\text{for}\quad 0\leq\patchidx<\npatches\}\; ,
\end{align*}
commute with the expected estimates \eqref{eq.diriestimate} and \eqref{eq.neumestimate}. For the spatial approximation error in elastodynamics, we refer to the same error estimates. There, we approximate functions in vector-valued Sobolev spaces $\myvec{f}_0,\myvec{f}_1\in({\hilbert}^{1/2}(\boundary))^3$ where assumptions on their regularities are made for each component.

\section{Numerical examples} \label{sec.examples}
The following numerical examples verify the correct implementation of the proposed methodology through a systematic validation approach. First, a convergence study in the Laplace domain is presented to show that the kernels in the CQM are correctly evaluated. Second, we examine the main contribution of this work by studying time domain results and their convergence rates. This validation proceeds in two stages: initially, the indirect formulations in \eqref{eq:diri_indirect} and \eqref{eq:neum_indirect} are used to compute the analytically known density functions of a scattering sphere. Subsequently, the Cauchy data of the same scatterer with mixed boundary conditions are studied to show the behavior of the direct formulation. The method is implemented in HyENA, our in-house code~\cite{hyena}, that utilizes some geometry preparation routines from the IGA-BEM code BEMBEL~\cite{DOLZ2020100476}.

\begin{figure}[htb]
  \centering
  \begin{minipage}{.49\textwidth}
    \centering
    \includegraphics[width=.8\textwidth]{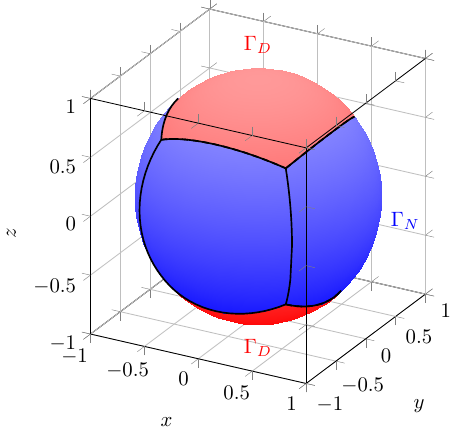}
    \captionof{figure}{The sphere constructed by six patches.}
    \label{fig.sphere2}
  \end{minipage}\hfill
  \begin{minipage}{.49\textwidth}
    \centering
    \captionof{table}{Mesh details.}
    \begin{tabular}{llll} \toprule
    level &  \#elements & $\nsteps$ & $\timestep$ [s] \\ \midrule
    0 &  6          & 2            & 5               \\
    1 &  24         & 4            & 2.5             \\
    2 &  96         & 8            & 1.25            \\
    3 &  384        & 16           & 0.625           \\
    4 &  1536       & 32           & 0.3125          \\ \bottomrule
    \end{tabular}
    \label{tab.mesh}
  \end{minipage}
\end{figure}

The unit sphere in $\Rn{3}$ embodied by six fourth order Bézier patches as the boundary $\boundary$, see \Cref{fig.sphere2},
serves as the considered manifold. We choose the material parameters in a way that we have the wave speeds of $\unitfrac[1]{m}{s}$ for the respective pressure and compression waves. The error analysis is layed out based on an uniform $h$-refinement in space and an uniform $\timestep$-refinement of the timeline, both refinements are executed simultaneously. The represented error is measured by a combined relative error in space and time
\begin{align}\label{eq.spacetime_{\elemidx}rror}
  \funnorm{{f}(\xb,\timevar) - \approximated{f}(\xb,\timevar)}{\discreteltwo{\boundary\times[0,\timeend]}} := \left(\timestep\sum\limits_{\timestepidx=0}^{\nsteps}\frac{\funnorm{{f}(\xb,\timestepidx\timestep) - \approximated{f}(\xb,\timestepidx\timestep)}{\Ltwo{\boundary}}^2}{\funnorm{{f}(\xb,\timestepidx\timestep)}{\Ltwo{\boundary}}^2}\right)^{1/2} \;.
 \end{align}
For the elliptic results only the spatial $L^2$-error will be presented.

The application of mixed boundary conditions implies the decomposition of the geometry into Dirichlet and Neumann patches, see \Cref{fig.sphere2}.
The red part of the boundary is $\boundary_D$ and the remaining blue part $\boundary_N$. 
In the following examples we refer to the source point as $\myvec{s}:=[1.5,1.5,1.5]$, the distance vector as $\myvec{\dist}:=\myvec{s}-\xb$ and address its Eucleadian norm by $\dist:=\left\|\myvec{\dist}\right\|$. For the considered tests in the time domain the CQM performes based on 3-stage and 5-stage Radau IIA methods with the respective stage orders $\rkstageorder=3$ and $\rkstageorder=5$ and the respective classical orders $\rkorder=5$ and $\rkorder=9$. As the initial discretization of the timeline we set $\nsteps=2$ time steps of size $\timestep=5\ s$. Details of the mesh in space and time can be studied in \Cref{tab.mesh}. The ploynomial degrees for the spatial approximation of the Dirichlet data is denoted as $\diridegree$ and for the Neumann data $\neumdegree=\diridegree-1$, respectively.

\subsection{Results in the Laplace domain}
As mentioned above, first, the kernel implementation is tested. Hence, the respective elliptic problems, i.e., the steady state solution at a distinct complex frequency $\laplacevar=1+i$ is computed.

\begin{figure}[H]
  \subcaptionbox{\label{fig.convergence_helmholtzDiri}Dirichlet data}{
    \includegraphics[width=.48\textwidth]{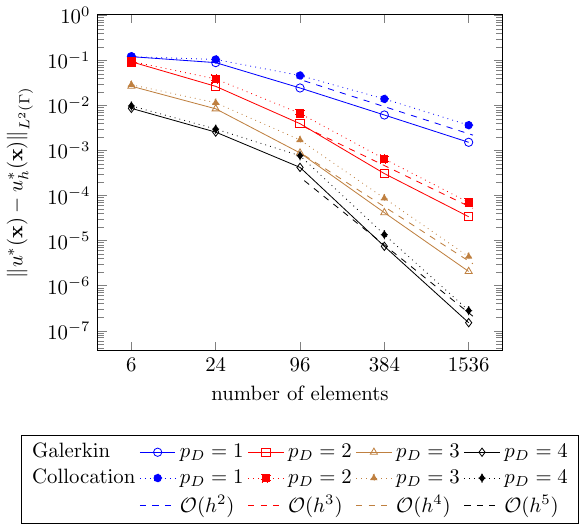}
  }   \hfill
  \subcaptionbox{\label{fig.convergence_helmholtzNeum}Neumann data}{
    \includegraphics[width=.48\textwidth]{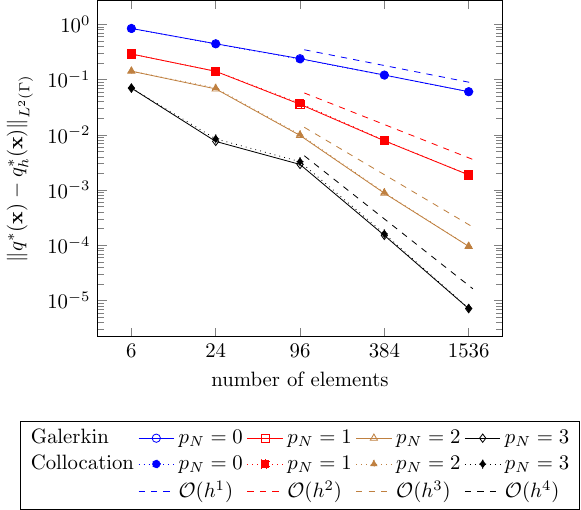}
  }
  \caption{Convergence plots of the Laplace domain solution in acoustics computed on the refinement levels $0,\hdots,4$.
  }
  \label{fig.convergence_helmholtz}
\end{figure}

\paragraph{Acoustics}
We investigate the approximation quality of the solution in the Laplace domain in a $\Ltwo{\boundary}$ error norm. The approximated solution of the Laplace domain fundamental solution is computed for the Laplace transformed set of equations corresponding to \eqref{eq.collo_bie} and \eqref{eq:bie} for the acoustic equation \eqref{eq.acoustics1}. The given boundary conditions are the fundamental solutions due to a source at $\myvec{s}$. Hence, the solution of the unknown data is as well given by the fundamental solutions. In \Cref{fig.convergence_helmholtz}, the error is given versus the number of elements on both boundary parts.
The expected optimal convergence rate of $\diridegree +1$ and $\neumdegree+1$ for the respective Dirichlet and Neumann data in the Laplace domain is observed. 

 \paragraph{Elastodynamics}
 The same study as above is performed for elastodynamics, i.e., the steady state response due to a dynamic loading governed by \eqref{eq.elastodynamics1}. The principal setup of the test is the same as above and the computed error is presented in \Cref{fig.convergence_{\elemidx}lastostatics}. 
 \begin{figure}[H]
   \subcaptionbox{\label{fig.convergence_elastostaticsDiri}Dirichlet data}{
     \includegraphics[width=.48\textwidth]{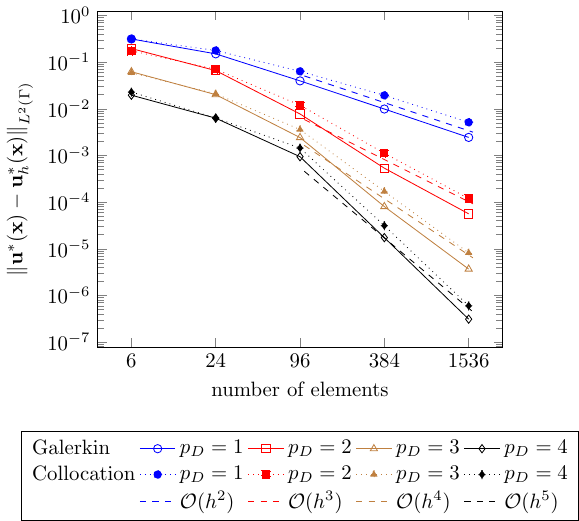}
   }   \hfill
   \subcaptionbox{\label{fig.convergence_{\elemidx}lastostaticsNeum}Neumann data}{
     \includegraphics[width=.48\textwidth]{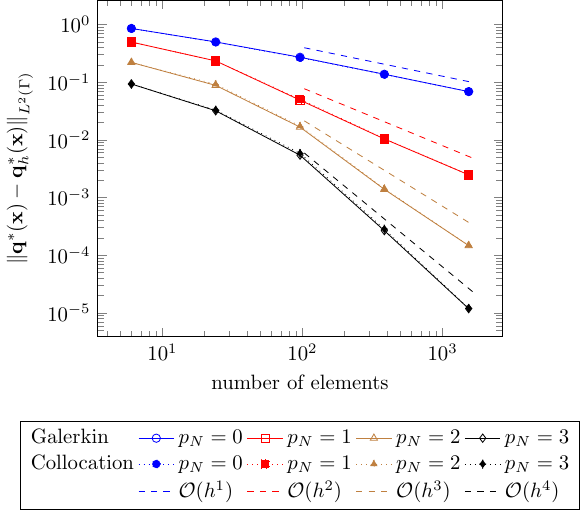}
   }
   \caption{Convergence plots of the Laplace domain solution in elastodynamics computed on the refinement levels $0,\hdots,4$.
   }
   \label{fig.convergence_{\elemidx}lastostatics}
 \end{figure}
 As well, the expected rate of convergence is obtained.
 
Both, studies show the correctness of the implementation of the IGA and the integration.
\subsection{Scattering sphere in the time domain}
After showing the correct IGA implementation, next, the time domain response is studied to show that the proposed method can achieve higher-order convergence rates.

 \subsubsection{Dirichlet and Neuman problem: Indirect formulation}
 The indirect method allows us to investigate the convergence rates of each boundary integral operator in \eqref{eq.operators1} and \eqref{eq.operators2} separately. For the excitation
 \begin{align} \label{eq:rhs}
   \timefunction(\xb,\timevar) = (\wavespeed\timevar)^\ell\exp(-\wavespeed\timevar) \sum\limits_{m=-n}^{n}\sh{n}{m}(\theta,\approxbasis)
 \end{align}
of a scattering sphere the density fields $\densityfirst$ and $\densitysecond$ on $\boundary$ are known analytically. In~\cite{veit2011numerical}, these solutions for Dirichlet and Neumann problems are presented for all of the formulations in \eqref{eq:diri_indirect} and \eqref{eq:neum_indirect}. For the Dirichlet problem the solutions can as well be found in~\cite{sauterveit13b}. To be precise, in the literature the solutions are given for arbitrary differential time functions. For the following results, however, we assume the time dependency as given in \eqref{eq:rhs}. The spatial dependency is modelled by spherical harmonics $\sh{n}{m}(\theta,\approxbasis)$, where in the following only the lowest order functions are used, i.e., $n=0$ is selected in the following tests. This is essentially a constant function over the sphere. The exponent  $\ell$  in \eqref{eq:rhs} governs the regularity of the time dependency. As required by the estimate \eqref{eq.energynormestimate}, $\ell-1$ derivatives have to be smooth at $t=0$. Hence, in the following examples, we choose $\ell=13$ such that the requirement $\ell>\max\{\rkorder+\mu+1,\mydegree_{\timevar},\rkstageorder+1\}$ is fulfilled for the considered different integral operators and RK methods.

\paragraph{Dirichlet problem}

\begin{figure}[hbt]
  \subcaptionbox{\label{fig:diri_ind_slp}Single layer approach}{
    \includegraphics[width=.48\textwidth]{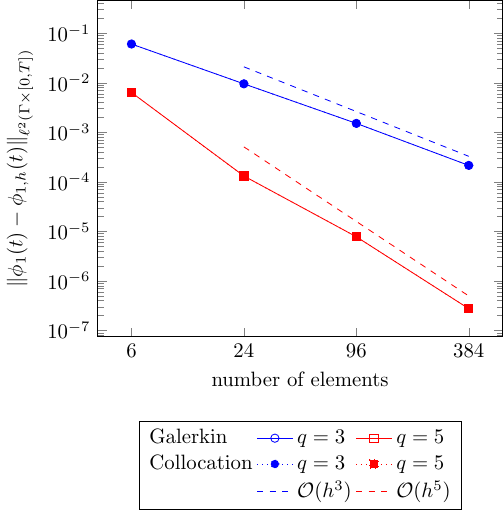}
  } \hfill
  \subcaptionbox{\label{fig:diri_ind_dlp}Double layer approach}{
    \includegraphics[width=.48\textwidth]{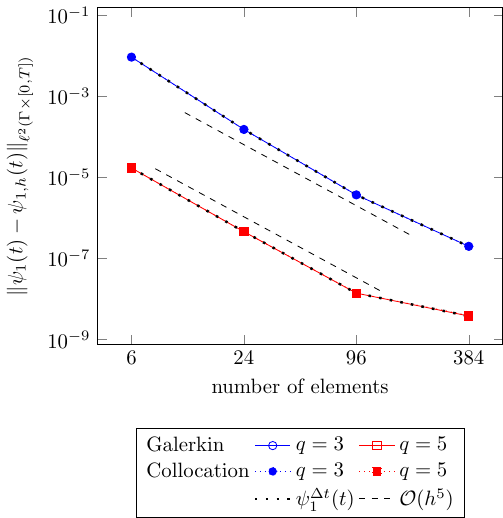}
  }
  \caption{Indirect formulation for the scattering sphere: Dirichlet problem. Convergence plots of the time domain solution computed on the refinement levels $0,\hdots,3$. In (b), the black dotted lines show the convergence of the CQM using the symbol of $\inverse{(\frac{1}{2} \densitysecond_1+\dlp*\densitysecond_1)}$.}
  \label{fig:diri_indirect}
\end{figure}

Both integral formulations in \eqref{eq:diri_indirect} can be used to compute the Dirichlet problem, i.e., a scattering sphere with sound soft boundary conditions. As discussed above the given Dirichlet datum is \eqref{eq:rhs}.  The analytical solution for $n=0$ reads for the single layer approach $\slp * \densityfirst_1 = \timefunction$
  \begin{equation*}
    \densityfirst_1(\timevar) = \frac{1}{\sqrt{\pi}}\sum\limits_{k=0}^{\lfloor {\timevar}/{2}\rfloor}\dot\timefunction(\timevar-2k) 
  \end{equation*}
  and for the double layer approach $\frac{1}{2} \densitysecond_1+\dlp*\densitysecond_1=\timefunction$
  \begin{align*}
    \densitysecond_{1}(\timevar) = \frac{1}{\sqrt{\pi}}\sum\limits_{k=0}^{\lfloor\timevar/2\rfloor}\sum\limits_{\ell=1}^{k+1}(-1)^{k} \int\limits_{2k}^{\timevar}c_{k\ell}(\tau-2k)^{k-\ell+1}\exp(\tau-2k)\dot\timefunction(\timevar-\tau)d\tau   && \text{with} &&  c_{k\ell}=\begin{pmatrix}k\\\ell-1 \end{pmatrix}\frac{2^{k-\ell+1}}{(k-\ell+1)!}\; .
  \end{align*}
  The error of both density functions is computed with \eqref{eq.spacetime_{\elemidx}rror} and displayed in \Cref{fig:diri_indirect}.
The expected convergence is $\min\{\rkorder,\rkstageorder+1-\mu+\nu, \mydegree+1\}$ as discussed above. In \cite[Lemma 7]{banjai2011runge}, it was proven that $\inverse{\freq{\slp}}$ is limited by $\mu=2$ and $\nu=1$, see \eqref{eq.cqm_estimates}. Hence, for a sufficient high spatial discretzation the convergence rate should be the stage order $\rkstageorder$. This can be observed in \Cref{fig:diri_ind_slp}. 
For the double layer approach no estimates are known for $\inverse{(\frac{1}{2}\idty+\freq{\dlp})}$. Thus, we introduce an error graph $\densitysecond_{1}^{\timestep}(t)$ of a reference solution (black, dotted line in \Cref{fig:diri_ind_dlp}) to further investigate the convergence behavior. This solution is obtained by using the analytical solution of the density in the Laplace domain $\densitysecond_{1}^*(\laplacevar)=2\laplacevar/(\laplacevar-1+(\laplacevar+1)\exp(-2\laplacevar))$, hence, the symbol of $\inverse{(\frac{1}{2} \densitysecond_1+\dlp)}$ within the CQM, i.e., a computation without using the BEM for the spatial variable.
Note that the reference and BEM solution match very well and that the plot in \Cref{fig:diri_ind_dlp} shows a higher convergence rate than stage order $\rkstageorder$. 

These results verify that the proposed method holds convergence rates of stage order $\rkstageorder$ for the solution to $\slp * \densityfirst_1 = \timefunction$ and also enables convergence rates of 5 for the solution to $\frac{1}{2} \densitysecond_1+\dlp*\densitysecond_1=\timefunction$. The spatial variable in this example is approximated by B-splines of degree $\mydegree=4$. Certainly, if a too low order $\mydegree$ for the spatial approximation is selected this order would dominate the convergence behavior.

\paragraph{Neumann  problem}

\begin{figure}[hbt]
  \subcaptionbox{\label{fig:neum_ind_slp}Single layer approach}{
    \includegraphics[width=.48\textwidth]{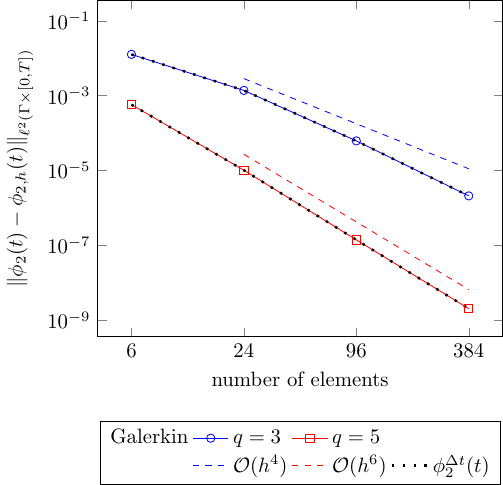} 
  } \hfill
  \subcaptionbox{\label{fig:neum_ind_dlp}Double layer approach}{
    \includegraphics[width=.48\textwidth]{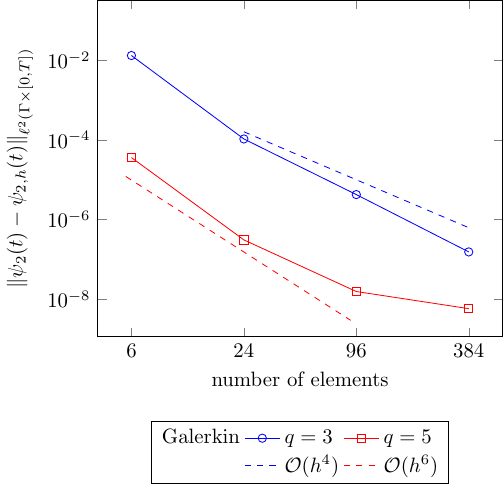}
  }
  \caption{Indirect formulation for the Scattering sphere: Neumann problem. Convergence plots of the time domain solution computed on the refinement levels $0,\hdots,3$. In (a), the black dotted lines show the convergence of the CQM using the symbol of $- \inverse{(\frac{1}{2} \densityfirst_2 - \adlp*\densityfirst_2)}$.}
  \label{fig:neum_indirect}
\end{figure}

As for the Dirichlet problem two approaches are available, the single layer and double layer approach listed in \eqref{eq:neum_indirect}. The Neumann problem describes a scattering sphere with sound hard boundary conditions. As discussed above the given Neumann datum is \eqref{eq:rhs}.  The analytical solution for $n=0$ reads for the single layer approach $- (\frac{1}{2} \densityfirst_2 - \adlp*\densityfirst_2) = \timefunction$
  \begin{equation*}
    \densityfirst_2(\timevar) = \frac{1}{\sqrt{\pi}} \left(-\sum\limits_{k=0}^{\lfloor {\timevar}/{2}\rfloor}\timefunction(\timevar-2k) + \sum\limits_{k=0}^{\lfloor {\timevar}/{2}\rfloor}\int\limits_{2k}^{\timevar}\exp(2k-\tau)\timefunction(\timevar-\tau)d\tau\right)
  \end{equation*}
  and for the double layer approach $- \hsop\densitysecond_2 = \timefunction$
  \begin{align*}
  \densitysecond_2(\timevar) = \frac{1}{\sqrt{\pi}}\left(-\int\limits_{0}^{\timevar}\timefunction(\timevar-\tau)\cosh(\tau)d\tau +  \sum\limits_{k=1}^{\lfloor\timevar/2\rfloor}\sum\limits_{\ell=1}^{k}(-1)^{k+1}\int\limits_{2k}^{\timevar}\frac{k-\ell+1}{2k}c_{k\ell}(\tau-2k)^{k-\ell+1}\exp(\tau-2k)\dot\timefunction(\timevar-\tau)d\tau\right)\; .
\end{align*}
The error of both density functions is computed with \eqref{eq.spacetime_{\elemidx}rror} and displayed in \Cref{fig:neum_indirect}.
Similar to the Dirichlet problem for the inverse of the adjoint double layer there are no values for $\mu$ and $\nu$ in \eqref{eq.kernel1} available. The plot shows the similar behavior to the Dirichlet case with a convergence rate of more than the stage order. It is even higher with respect to the spatial degree $\mydegree$. Again as comparison, the behavior of the solution $\densityfirst_{2}^{\timestep}(t)$  utilizing CQM based on the symbol of $- \inverse{(\frac{1}{2} \densityfirst_2 - \adlp*\densityfirst_2)}$, thus the exact solution in the Laplace domain $\densityfirst_{2}^{*}(\laplacevar)=-\laplacevar/(\sinh(\laplacevar)(\laplacevar+1)\exp(-\laplacevar))$ is displayed as the black dotted line in \Cref{fig:neum_ind_slp}. It cannot be distinguished from the BEM solution. In the double layer approach, the inverse hyper singular operator $\inverse{\freq{\hsop}}$ is used. Its estimate in \eqref{eq.cqm_estimates} gives $\mu=1,\,\nu=1$, which results in an expected convergence rate of  $\rkstageorder+1$. This convergence rate is observed in \Cref{fig:neum_ind_dlp}. Hence, again the proposed method gives the expected convergence rates such that it is indeed a higher-order method.
\newpage 
\subsubsection{Mixed problem: Direct formulation}
After presenting the behavior of all four boundary integral operators, a mixed problem is shown based on the formulations \eqref{eq.collo_bie} and \eqref{eq:bie}. There is no analysis available, however, it can expected that at least the behavior of the single layer potential will carry over to the mixed problem, i.e., it is expected that either the order of the spatial discretzation will dominate or the stage order. Next, we consider acoustics and elastodynamics, where the approximation order in time is fixed by using the 3-stage Radau IIA method as the Runge–Kutta scheme in the CQM, while the polynomial degree of the spatial approximation is varied. The considered manifold is again the sphere constructed by 6 patches as presented in \Cref{fig.sphere2}.

\begin{figure}[hbt]
  \subcaptionbox{\label{fig.convergence_acousticsPress}Pressure solution}{
    \includegraphics[width=.48\textwidth]{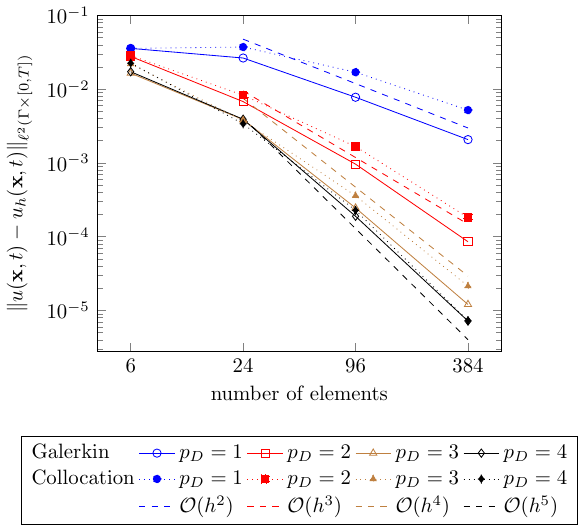}
  }   \hfill
  \subcaptionbox{\label{fig.convergence_acousticsFlux}Flux solution}{
    \includegraphics[width=.48\textwidth]{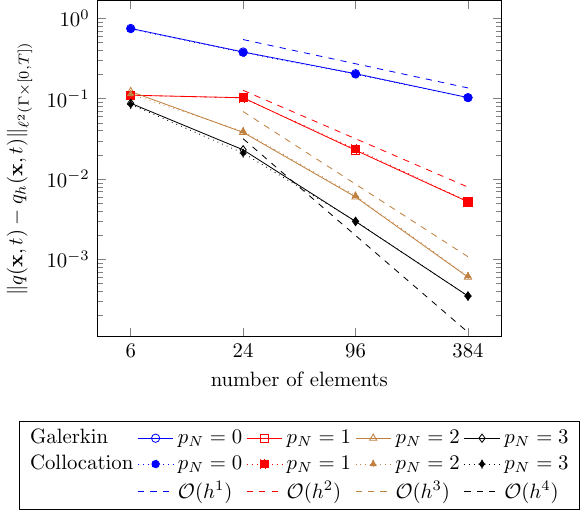}
  }
  \caption{Convergence plots of the time domain solution in acoustics computed on the refinement levels $0,\hdots,3$.}
  \label{fig.convergence_acoustics}
\end{figure}

\paragraph{Acoustics}
To obtain a reference solution, a bump function in time with an retarded argument $\timevar - \frac{\dist}{\wavespeed}$ is selected
\begin{equation}\label{eq.ansatz_timefunction}
  \pressure(\xb,\,\timevar)= \dist^{-1}\heaviside{\timevar - \frac{\dist}{\wavespeed}} \left(\timevar-\frac{\dist}{\wavespeed}\right)^\ell \exp\left(\dist-\wavespeed\right) \; .
\end{equation}
The exponent $\ell=9$ ensures sufficient regularity at $t=0$ and the Heaviside function $\heaviside \timevar$ ensures zero conditions for $\timevar<0$. For setting up a mixed problem as well, the analytical time solution of the acoustic flux is necessary. The normal derivative of the pressure field \eqref{eq.ansatz_timefunction} gives
\begin{equation*}
  \flux(\xb,\timevar)= \left(\frac{1}{2\pi\dist^3} + \frac{2(\dist-\wavespeed\timevar)}{\pi\dist^2} \right)\exp(-2(\dist-\wavespeed\timevar)) \langle \myvec{\dist},\myvec{n}\rangle \; .
\end{equation*}
Applying these boundary conditions with the source point $\myvec{s}$ allows us to compute the error defined in \eqref{eq.spacetime_{\elemidx}rror}. It is plotted versus an uniform  refinement in space and time in \Cref{fig.convergence_acoustics}.
In this test where we expect a convergence rate of at least stage order $\rkstageorder=3$ from the CQM, we can see in \Cref{fig.convergence_acoustics} that the convergence rates up to the order 3 are primarily influenced by the spatial spline approximations. Higher convergence rates are not feasable with the 3-stage Radau IIA method. It is observed that the convergence rate of the Dirichlet data (pressure) is $\diridegree+1$ and that of the Neumann data (flux) is $\neumdegree+1$. Hence, also for the mixed problem the proposed method behaves as expected and is indeed higher-order.

\paragraph{Elastodynamics}
To obtain analytical solutions for the three-dimensional computations in elastodynamics some kind of fundamental solutions are constructed (see,~\cite{kager2015fast,eringen1975elastodynamics}). They are no time domain fundamental solutions in the usual understanding as the selected excitation in time is given by the smooth function
\begin{equation}
  \text{F}( \dist,\timevar, \wavespeed_\alpha) = \exp\left(-a \left( t - \frac{r}{\wavespeed_\alpha} - a b \right)^2 \right) \; .
\end{equation}
This is a smooth pulse, where the parameters are chosen to be $a=0.1$ and $b=100$. Those values fit to the selected arbitrary material data with a compression wave speed of $c_1=\unitfrac[1]{m}{s}$ and the shear wave speed $c_2=\unitfrac[1/\sqrt{2}]{m}{s}$. Fundamental solutions in elastodynamics are tensors because not only the displacements are vectors but as well the direction of the applied force matters. To obtain the components of the sought solution $\displacement(\xb, \timevar)$ and $\stress(\xb, \timevar)$ a multiplication with a direction vector $\direction=\transpose{[1,1,1]}$ and the outward normal vector $\nvec{\xb}$ is necessary. This gives the known and unknown boundary data
\begin{equation}
  \displacement_k (\xb,\timevar) = \myvec{U}_{k \ell}(\myvec{\dist}, \timevar)\direction_\ell \qquad\text{and}\qquad \stress_k(\xb,\timevar)=\myvec{Q}_{k \ell m}(\myvec{\dist}, \timevar)\myvec{n}_\ell(\xb)\direction_m \; .
\end{equation}
The expressions for the fundamental tensors can be found in \Cref{app.anaElasto}.

\begin{figure}[htb]
  \subcaptionbox{\label{fig.convergence_elastodynamicsDisp}Displacement solution}{
    \includegraphics[width=.48\textwidth]{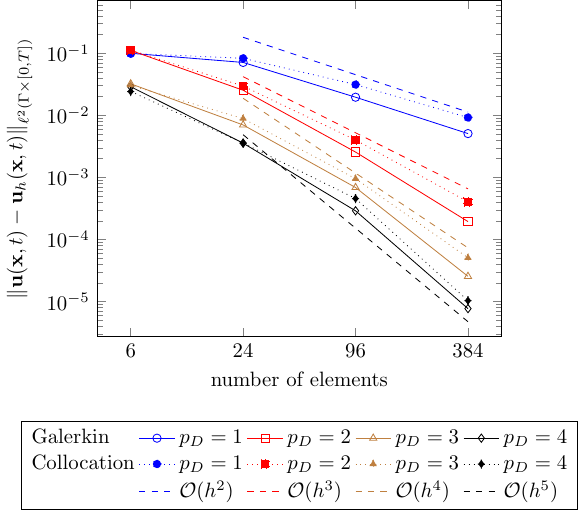}
  }   \hfill
  \subcaptionbox{\label{fig.convergence_elastodynamicsTrac}Traction solution}{
    \includegraphics[width=.48\textwidth]{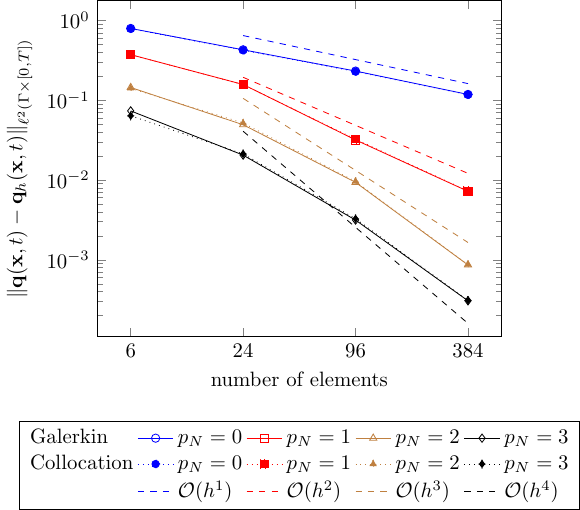}
  }
  \caption{Convergence plots of the time domain solution in elastodynamics on the refinement levels $0,\hdots,3$.}
  \label{fig.convergence_elastodynamics}
\end{figure}

The mixed error in space and time is computed based on the magnitutes of $\displacement(\xb, \timevar)$ and $\stress(\xb, \timevar)$, respectively. In \Cref{fig.convergence_elastodynamics}, we can see a similar convergence behavior as in the case of acoustics shown above. While choosing spline approximations of degree $\diridegree=1,\hdots,4$ and $\neumdegree=0,\hdots,3$ the convergence rate of the Dirichlet data (deformation) and that of the Neumann data (normal surface stresses) is respectively restricted by $\diridegree+1$ and $\neumdegree+1$ until the convergence rate of the stage order $\rkstageorder=3$ of the 3-stage Radau IIA method is reached.

\newpage
\section{Conclusions}
This work presents a higher-order method for solving time-dependent partial differential equations arising in acoustics and elastodynamics. This method discretizes the timeline using the convolution quadrature method (CQM) based on a multi-stage Runge-Kutta scheme. The resulting elliptic problems in the CQM are solved with isogeometric analysis (IGA), where the Laplace-domain solutions are approximated with splines of higher polynomial degree. To facilitate implementation, we outline the necessary basis transformations to localize the basis functions on the elements, allowing the use of preexisting integration routines. While our primary objective was developing a higher-order method for mixed boundary value problems, the exact parameterization of spheres through rational spline mappings allowed us to study individual boundary operators separately using indirect formulations with known analytical solutions. The analysis on the CQM and the IGA provide optimal convergence rates separately, in time and space. Numerical examples confirm that this behavior manifests in the overall space–time convergence, with the convergence rate limited by the smaller of the two.

Maintaining the optimal rate of convergence requires accurate evaluation of the integrals appearing in the boundary operators. These integrals pose two major challenges: they involve singular integrands as $\dist \to 0$, and they become highly oscillatory when the complex frequencies have large imaginary components. To address these difficulties, quadrature rules with $20 \times 20$ Gauss points have been employed, though this leads to significant computational expense. Therefore, for efficient integration, the use of frequency-dependent quadrature rules is highly recommended. Adaptive quadrature rules for integrals in boundary integral equations are given in \cite{harbrecht2006wavelet} and work well, particularly for functions of the type $\frac{1}{\dist}$. However, for oscillatory functions, alternative quadrature schemes need to be considered, as studied in \cite{iserles2004quadrature} and \cite{huybrechs2007construction}.

\paragraph*{Acknowledgement}
This work is supported by the joint DFG/FWF Collaborative Research Centre CREATOR (DFG: Project-ID 492661287/TRR 361; FWF: 10.55776/F90) at TU Darmstadt, TU Graz and JKU Linz.

\appendix
\section{Analytical elastodynamic solution in time domain} \label{app.anaElasto}
The analytic reference solutions for elastodynamics describe the vectorial wave propagation of an unit impuls originated at $\myvec{s}$ applied to the full space. We refer to \cite{kager2015fast,eringen1975elastodynamics} and introduce the second order displacement tensor for $k,\ell=1,2,3$ as
\begin{equation}
  \myvec{U}_{k \ell}(\myvec{\dist}, \timevar) = \frac{1}{4\pi \rho r} \left[ 
    \myvec{f}^0_{k \ell}(\myvec{\dist}) \, {I}(\dist, \timevar) 
    + \myvec{f}^1_{k \ell}(\myvec{\dist}) \, \text{F}(r,\timevar, c_1) 
    + \myvec{f}^2_{k \ell}(\myvec{\dist}) \, \text{F}(r,\timevar, c_2)
  \right]
\end{equation}
with the functions $\myvec{f}^0_{k \ell}(\myvec{\dist}) = \frac{1}{r} \left( 3 \distvec_k \distvec_\ell - \kronecker_{k \ell} \right)$, $\myvec{f}^1_{k \ell}(\myvec{\dist}) = \frac{\distvec_k \distvec_\ell}{r c_1^2}$, and $\myvec{f}^2_{k \ell}(\myvec{\dist}) = \frac{1}{r c_2^2} \left(\kronecker_{k \ell}  - \distvec_k \distvec_\ell \right)$. Note that $\mathbf{\dist} = \xb-\yb$ and $r=\|\xb-\yb\|$ holds.  Further, $F(r,\timevar, \wavespeed_\alpha)$ needs to be twice differentiable with respect to time and  is chosen as
\begin{equation}
  \text{F}( \dist,\timevar, \wavespeed_\alpha) = \exp\left(-a \left( t - \frac{r}{\wavespeed_\alpha} - a b \right)^2 \right)
\end{equation}
with constant parameters $a,\,b$. In this example the parameters are chosen to be $a=0.1$ and $b=100$. Finally, the integral term reads
\begin{align*}
  {I}(\dist,\timevar) &= \int\nolimits_{c_{1}^{-1}}^{c_{2}^{-1}}\lambda \text{F}(\timevar-\lambda r) d\lambda \\
                      & =
                        \frac{1}{2 a \dist^2} \left[
                        \exp\left(-q_{1}^2(\dist,\timevar)\right) 
                        - \exp\left(-q_{2}^2(\dist,\timevar)\right)+ 
                        \sqrt{a\pi}(ab-t)\left( E\left(q_{1}(\dist,\timevar)\right)
                        - E\left(q_{2}(\dist,\timevar)\right)
                        \right)\right]
\end{align*}
with the error function $E(x, \timevar) = \frac{2}{\sqrt{\pi}}\int\limits_{0}^{x}\exp(-t^2)dt$ and $q_{\alpha}(\dist,\timevar) =  \frac{\sqrt{a}}{c_{\alpha}}(a b c_{\alpha} + r - c_{\alpha} t)$. The reference solution for the traction is then defined as a third order tensor and reads for $k,\ell,m=1,2,3$ as
\begin{align}
  \begin{split}
    \myvec{Q}_{k \ell m}(\myvec{\dist},\timevar) =& \frac{\matdensity}{4\pi}\left[ \myvec{G}^0_{k \ell m}(\myvec{\dist}){I}(\dist,\timevar)+\myvec{G}^1_{k \ell m}(\myvec{\dist})\left(F\left(\dist, \timevar, \wavespeed_2\right) - \left(\frac{\wavespeed_2}{\wavespeed_1}\right)^2 F\left(\dist, \timevar, \wavespeed_1\right)\right) \right. \\
    &\left. +\myvec{G}^2_{k \ell m}(\myvec{\dist})\left(\dot{F}\left(\dist, \timevar, \wavespeed_2\right) - \left(\frac{\wavespeed_2}{\wavespeed_1}\right)^3 \dot{F}\left(\dist, \timevar, \wavespeed_1\right)\right)+\myvec{G}^3_{k \ell m}(\myvec{\dist})\left(F\left(\dist, \timevar, \wavespeed_1\right) + \frac{\dist}{\wavespeed_1} \dot{F}\left(\dist, \timevar, \wavespeed_1\right)\right) \right.\\
    &\left. +\myvec{G}^4_{k \ell m}(\myvec{\dist})\left(F\left(\dist, \timevar, \wavespeed_2\right) - \frac{\dist}{\wavespeed_2} \dot{F}\left(\dist, \timevar, \wavespeed_2\right)\right)\right]
  \end{split}
\end{align}
with
\begin{align*}
  \myvec{G}^0_{k \ell m}(\myvec{\dist}) & = -\frac{6\wavespeed_2^2}{\dist^2}\left(\frac{5\distvec_k\distvec_\ell\distvec_m}{\dist} - \frac{\kronecker_{k \ell}\distvec_m + \kronecker_{k m}\distvec_\ell + \kronecker_{\ell m}\distvec_k}{\dist} \right),\\
  \myvec{G}^1_{k \ell m}(\myvec{\dist}) & = \frac{2}{\dist^2}\left(\frac{6\distvec_k\distvec_\ell\distvec_m}{\dist} - \frac{\kronecker_{k \ell}\distvec_m + \kronecker_{k m}\distvec_\ell + \kronecker_{\ell m}\distvec_k}{\dist} \right),\\
  \myvec{G}^2_{k \ell m}(\myvec{\dist}) & = \frac{2\distvec_k\distvec_\ell\distvec_m}{\dist^4\wavespeed_2},\\
  \myvec{G}^3_{k \ell m}(\myvec{\dist}) & = -\frac{\kronecker_{k \ell}\distvec_m}{\dist^3}\left[ 1-2\left(\frac{\wavespeed_2}{\wavespeed_1}\right)^2\right],\\
  \myvec{G}^4_{k \ell m}(\myvec{\dist}) & = -\frac{1}{\dist^2}\left(\frac{\kronecker_{k m}\distvec_\ell}{\dist}+\frac{\kronecker_{\ell m}\distvec_k}{\dist}\right).\\
\end{align*}

%
\bibliography{mybibfile}
\bibliographystyle{unsrtnat}
\end{document}